%% file: main_v3.tex
\documentclass[twocolumn,amsthm]{autart}%

\usepackage{ifpdf}

\usepackage[round]{natbib}%

\usepackage{wrapfig}
\usepackage{graphicx}
\usepackage{subcaption}
\graphicspath{{figs/}}

\usepackage{parskip}

\usepackage{mathtools,amsmath}

\usepackage{amsfonts,amssymb}
\usepackage{bm} %
\usepackage{mathrsfs}  %
\usepackage{cases}  %
\usepackage{amsmath}
\usepackage{comment}
\usepackage{tabularx}
\newcommand{\multiline}[1]{%
  \begin{tabularx}{\dimexpr\linewidth-\ALG@thistlm}[t]{@{}X@{}}
    #1
  \end{tabularx}
}
\usepackage{diagbox}
\newcolumntype{K}[1]{>{\centering\arraybackslash}p{#1}}

\usepackage{indentfirst}
\usepackage[per-mode=symbol]{siunitx}

\usepackage[svgnames, rgb]{xcolor}  %

\usepackage{tikz}
\usetikzlibrary{snakes,arrows,shapes,positioning}

\usepackage{pgfplots}
\pgfplotsset{compat=1.15}
\usepackage{mathrsfs}
\usetikzlibrary{arrows}

\usepackage{multirow}

\usepackage{algorithm}
\usepackage{algpseudocode}  %

\usepackage{cite}  %
\usepackage{array}  %
\usepackage{booktabs}           %

\usepackage{amsthm}
\theoremstyle{definition}
\newtheorem{assumption}{Assumption}
\newtheorem{definition}{Definition}
\newtheorem{example}{Example}

\newtheorem{remark}{Remark}

\theoremstyle{plain}
\newtheorem{proposition}{Proposition}

\usepackage[extramath,probability,reviewmark]{mylatexdefs}

\newcommand{\hilitediff}[1]{#1}

\input{defs}

\ifpdf
  \usepackage[pdftex]{hyperref}
\else
  \usepackage[dvips]{hyperref}  %
\fi

\begin{document}
\begin{frontmatter}
\title{Distributed Mixed-Integer Quadratic Programming for Mixed-Traffic Intersection Control} %
\thanks[footnoteinfo]{This research was supported in part by NSF under Grants CNS-2149520, CMMI-2348381, IIS-2415478, and in part by Mathworks.}
\thanks[footnoteinfo]{This work was conducted while Viet-Anh Le was with the Department of Mechanical Engineering, University of Delaware, Newark, DE 19716, USA, and the Systems Engineering Program, Cornell University, Ithaca, NY 14850, USA.}

\author[UPENN]{Viet-Anh Le}\ead{vietanh@engineering.upenn.edu},
\author[SY,CEE]{Andreas A. Malikopoulos}\ead{amaliko@cornell.edu}

\address[UPENN]{University of Pennsylvania, Philadelphia, PA 19104 USA}
\address[SY]{Systems Engineering Program, Cornell University, Ithaca, NY 14850 USA}
\address[CEE]{School of Civil and Environmental Engineering, Cornell University, Ithaca, NY 14853 USA}

\begin{keyword}                           %
Connected and automated vehicles, traffic signal control, mixed traffic, mixed-integer quadratic programming.
\end{keyword}                             %

\begin{abstract}
In this paper, we develop a distributed solution framework for coordinating traffic light systems and connected automated vehicles (CAVs) in mixed-traffic intersections.
We formulate a mixed-integer quadratic program (MIQP) aimed at optimizing the coordination of traffic light systems and CAVs, thereby fully capitalizing on the advantages of CAV integration under conditions of high penetration rates. 
To effectively solve the challenging multi-agent MIQP approximately, we employ proximal ADMM for solving the convex relaxation of the MIQP while systematically tightening the constraint coefficients using a finite-grid tightening rule to uphold integrality requirements.
\hilitediff{The obtained relaxed solutions are corrected using a confidence-guided binary recovery step, which fixes high-confidence binary variables and solves a substantially reduced MIQP over the remaining uncertain binaries to improve the feasibility of the final mixed-integer solution.}
The performance of our control framework and the efficacy of the distributed algorithm are rigorously validated through a series of simulations conducted across varying penetration rates and traffic volumes.

\end{abstract}

\end{frontmatter}

\section{Introduction}

\subsection{Motivation}

Coordination of connected autonomous vehicles (CAVs) has shown promise in improving various traffic metrics, including energy consumption, greenhouse gas emissions, and travel time while ensuring safety, as demonstrated in several studies; \eg \citet{Malikopoulos2020,chalaki2020TCST,katriniok2022fully}.
However, a recent forecast indicates that by $2060$, the CAV penetration rate may still not exceed $90\%$ \citep{matin2024public}.
Therefore, addressing the planning and control of CAVs in mixed traffic, given the coexistence of human-driven vehicles (HDVs), is a crucial research focus and has gained significant attention in recent years \citep{le2024stochastic} for a comprehensive review.
Our early work on this topic focused on problems in unsignalized mixed-traffic scenarios with a single conflict point, such as merging at roadways \citep{Le2023CDC,le2024stochastic}, or single-lane intersections \citep{Le2022CDC,Le2023ACC}.
In more complex scenarios, such as multi-lane intersections, coordinating CAVs together with multiple HDVs without traffic signals may be unrealistic.
On the other hand, intelligent traffic signal control can exert a certain level of control over HDVs at intersections and is effective in ensuring traffic safety and alleviating congestion in regular traffic consisting entirely of HDVs \citep{varaiya2013max}.
Therefore, several recent works have explored mixed-traffic settings where traffic-signal optimization is combined with CAV coordination \citep[e.g.,][]{kamal2019development,du2021coupled,niroumand2023,ghosh2022traffic}. 
These studies highlight the potential benefits of leveraging both infrastructure-based control (signals) and vehicle-based control (CAV trajectories) in the presence of HDVs.
In particular, when CAV penetration rates are low, optimizing traffic signals is more critical to ensuring conflict-free vehicle maneuvers and improving traffic flow, while CAV trajectory optimization can still be considered to optimize energy consumption.
Meanwhile, under the high penetration rates of CAVs, their coordination can fully exploit the benefits of avoiding stop-and-go driving behavior.

\subsection{Literature Review}

\subsubsection{Traffic light and CAV coordination}

Combining traffic signal control with CAV coordination has emerged as a promising approach for managing mixed-traffic intersections \citep{li2023survey}.
The current state-of-the-art methods for the coordination of traffic light systems and CAVs in mixed-traffic intersections can be classified into three main categories: (1) \emph{reinforcement learning}, (2) \emph{bi-level optimization}, and (3) \emph{joint optimization}.

Reinforcement learning (RL) methods aim to train control policies that optimize a specified reward function using traffic simulators. 
Within this line of research, \citet{guo2023cotv} employed a multi-agent deep RL framework in which HDV behavior was modeled using the intelligent driver model (IDM), while \citet{song2021traffic} enhanced deep Q-network performance through policy transfer across similar traffic scenarios. 
However, these RL-based approaches generally face training instability in large-scale settings and typically lack the ability to ensure real-time safety.

Optimization-based approaches, including bi-level and joint optimization, explicitly enforce safety-related constraints and are therefore well-suited for dealing with uncertain HDV behavior. 
Bi-level optimization approaches separate traffic signal optimization from CAV trajectory planning: traffic signal optimization is solved using an approximate traffic model, followed by CAV trajectory optimization. 
For example, \citet{kamal2019development} used a receding horizon control framework to minimize approximate total crossing times using optimal traffic signals, while \citet{du2021coupled} proposed an upper-level signal-timing optimization that minimizes delay and balances phase durations, with the resulting signal plan used to generate CAV eco-driving trajectories via a trigonometric function-based arrival-time algorithm. 
These works simplify the optimization problem by separating signal control and CAV planning, but the resulting solutions depend heavily on the accuracy of the upper-level traffic model.

A more tightly integrated strategy is offered by joint optimization approaches, in which traffic-signal timings and CAV trajectories are optimized simultaneously.
Within this category, \citet{niroumand2023} proposed a white phase that uses CAVs as mobile traffic controllers, which negotiate the right-of-way to lead a group of trailing HDVs and formulated the joint signal timing and trajectory optimization problem as a mixed-integer nonlinear program (MINLP).
\citet{ghosh2022traffic} formulated an optimization problem considering a linear combination of energy cost and travel time, maximizing the velocity at which vehicles can enter the intersection, and minimizing the overall time taken for all vehicles to cross the intersection.
\citet{tajalli2021traffic} formulated a MINLP to maximize the distance traveled by each vehicle and minimize acceleration while optimizing the signal timing parameters through a cycle- and phase-free plan.
\citet{ravikumar2021mixed} formulated a mixed-integer linear program (MILP) for scheduling and coordination of CAVs in an interconnected network of conflict zones.
\citet{firoozi2022coordination} improved upon the work in \citet{ravikumar2021mixed} by introducing a switching dynamic model based on logic conditions for HDVs.
\citet{suriyarachchi2023optimization} presented a MILP formulation for joint vehicle and intersection coordination and developed a decentralized implementation by decoupling the MILP.
While these joint optimization approaches offer more direct coordination between signal decisions and vehicle trajectories, they generally lead to large mixed-integer problems, which pose computational challenges for real-time implementation.

\subsubsection{Distributed Algorithms for MICPs}

One of the main challenges for optimization-based frameworks for joint traffic signal control and CAV trajectory planning, or receding horizon motion planning in general, is real-time practicality, as these approaches require solving multi-agent mixed-integer optimization problems.
Thus, in this section, we focus on the algorithms developed to date that aim to solve multi-agent mixed-integer optimization.

Reducing computation time for finding a solution to multi-agent mixed-integer optimization can be achieved using decomposition-based distributed algorithms.
Several works \citep{vujanic2016decomposition,falsone2019decentralized,falsone2018distributed,camisa2018primal,camisa2021distributed,testa2019distributed} propose dual or primal decomposition, constraint tightening, or cutting-plane exchange to obtain feasible distributed solutions. 
However, the above papers are only applicable to MILPs since they rely on the Shapley-Folkman lemma \citep{bertsekas2014constrained}, which cannot be extended to a mixed-integer quadratic program (MIQP).
Moreover, the algorithms typically require the total number of subsystems (agents) to be much greater than the number of coupling constraints.

On the other hand, research on distributed algorithms for MIQPs is still limited but has received increasing attention.
\citet{liu2021distributed} proposed a three-stage algorithm to solve the MIQP problem; however, the algorithm requires an initial feasible guess of the solution.
\citet{liu2022distributed} proposed a two-stage alternating direction method of multipliers (ADMM) algorithm to solve MIQP by selecting penalty weights that satisfy certain conditions.
\citet{sun2018distributed} decoupled the large-scale problem into subproblems, where each local update is a convex projection subproblem except for one handling the integrality constraint. A distributed projected subgradient algorithm is then used.
Though these methods exploit distributed computation to reduce solving time, their implementation still involves solving mixed-integer subproblems.
\citet{klostermeier2024numerical} presented a dual decomposition-based distributed optimization method for finding a lower bound for the optimal solution in a branch-and-cut algorithm.
The use of ADMM algorithms as a heuristic for solving MIQP with relaxation was also explored in \citet{takapoui2020simple}, \citet{stellato2018embedded}, and \citet{alavian2017improving}.
However, since MIQP is a nonconvex problem, ADMM may fail to converge or may converge to an incorrect solution, as mentioned in \citet{takapoui2020simple}, even for a simple problem like integer linear programming.

\subsection{Our Contributions}

In our prior work \citet{le2024distributed}, we developed a formulation that leverages both CAV coordination at higher penetration rates and traffic-light management at lower penetration rates. 
Compared with related work, we incorporated mixed-integer lateral constraints to enable coordinated CAV crossing without relying solely on traffic lights, which introduced additional binary variables beyond those representing traffic-light states.
To solve the resulting multi-agent MIQP in a receding-horizon setting, we developed a penalization-enhanced maximum block improvement algorithm that is faster than centralized optimization with \texttt{GUROBI} \citep{gurobi}. 
However, the local subproblems remain MIQPs, and only one agent updates its solution at each iteration, leading to more iterations as the number of agents increases. 
To reduce complexity, our prior formulation used single binaries over the control horizon for lateral collision avoidance, which limits CAV coordination flexibility.

In this paper, we develop a simple and computationally efficient distributed algorithm for approximate solutions of the MIQP arising from joint traffic-signal control and CAV coordination in mixed traffic. 
Compared with \citet{le2024distributed}, we refine the formulation by allowing the lateral collision-avoidance binaries to vary over the control horizon, enabling more flexible CAV coordination. 
Since this significantly increases the number of binary variables, we propose a distributed optimization algorithm based on proximal ADMM and sequential constraint tightening. 
\hilitediff{We first solve the convex relaxation using parallel proximal ADMM, then promote integrality using a finite-grid tightening rule for the coefficients of big-$M$ constraints, which are commonly used in mixed-integer motion planning \citep{alrifaee2014centralized}. 
The obtained relaxed solution is corrected by a confidence-guided binary recovery step, which fixes high-confidence binaries and solves a reduced MIQP over the remaining uncertain binaries.} 
The resulting method efficiently obtains high-quality mixed-integer solutions while improving the feasibility of the recovered solution.

\subsection{Organization}
The remainder of this article is organized as follows.
Section~\ref{sec:prb} formulates an MIQP problem for joint traffic signal control and CAV coordination in mixed-traffic intersections.
Section~\ref{sec:tightening} and Section~\ref{sec:admm} present the sequential constraint tightening technique and the proximal ADMM algorithm for solving the multi-agent MIQP, respectively.
Section~\ref{sec:theo} analyzes the theoretical properties of the proposed algorithm.
Section~\ref{sec:sim} shows the simulation results to validate the effectiveness of the
proposed control framework.
Finally, we provide concluding remarks in Section~\ref{sec:conc}.

\section{Problem Formulation}
\label{sec:prb}

In this section, we present the MIQP formulation for the joint optimization problem of traffic light control and CAV coordination that was developed in \citet{le2024distributed} with an improvement designed to enhance CAV coordination.

\subsection{Modeling Framework}
  
\begin{figure}
\centering
\includegraphics[width=0.5\textwidth, trim=80 80 40 80, clip]{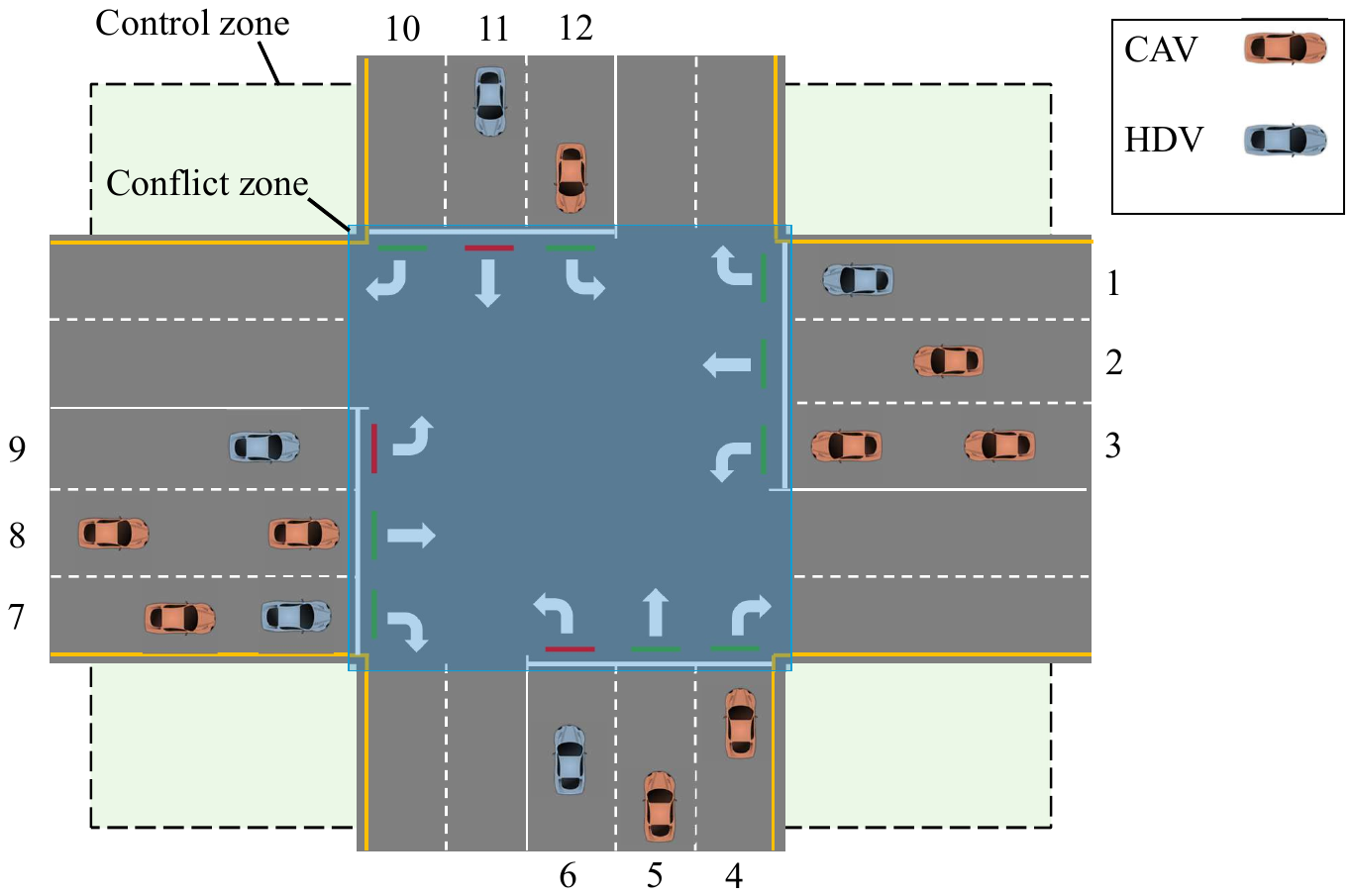}
\caption{An intersection scenario in mixed traffic with 12 lanes, including separate lanes for right turns, straight-through traffic, and left turns. The control zone and conflict zone are represented by the green and blue regions, respectively.
}
\label{fig:scenario}
\end{figure}

We focus on an isolated intersection, as depicted in Fig. \ref{fig:scenario}, considering dedicated lanes for right turns, through traffic, and left turns.
A \emph{control zone} is defined around the intersection, where CAVs and traffic light controllers (TLCs) can communicate and be managed within the proposed framework.
Moreover, we define a \emph{conflict zone} as the region where a lateral collision may occur.
To facilitate the implementation of the proposed control framework, we consider the following assumptions:

\begin{assumption}
\label{assp:1}
A central coordinator is available that stores information about the intersection’s geometric parameters and traffic light status, in addition to the real-time positions and speeds of the vehicles traveling within the control zone.
The central coordinator is also responsible for computing the control framework and communicating the optimal control decisions to the traffic lights and CAVs.
\end{assumption}

\begin{assumption}
\label{assp:2}
Lane changes are not permitted within the control zone.
\end{assumption}

Although the scenario in Fig. \ref{fig:scenario} includes 12 lanes, right-turn lanes can be treated separately since they do not conflict with other lanes, thus reducing the problem to 8 relevant lanes.
Next, we introduce several definitions for the entities involved in the intersection scenario.

\begin{definition}
(Lanes) Let \lane{l} be the $l$-th lane in the scenario and $\LLL$ be the set of all lanes' indices.
Each \lane{l} is equipped with a traffic light controller denoted by \TLC{l}.
We set each lane's origin location at the control zone's entry and let $\psi_l \in \RR$ and $\phi_l \in \RR$ be the positions of the conflict zone's entry and exit points along \lane{l}.
Note that $\psi_l$ can also be the position of the stop line of \lane{l}.
\end{definition}

\begin{definition}
(Vehicles) Let a tuple $(l,i)$ be the index of the $i$-th vehicle traveling in \lane{l}.
For each \lane{l}, let $\CCC_l(k)$ and $\HHH_l(k)$ be the sets of CAVs and HDVs in \lane{l} at any time step $k$.
\end{definition}

\begin{table*}[t]
\centering
\caption{Notation summary for traffic light variables, CAV dynamics, and model parameters.}
\label{tab:notation}
\begin{tabular}{p{4cm} p{12cm}}
\toprule
\textbf{Notations} & \textbf{Description} \\
\midrule
$\Delta T$ & Sampling period \\[2pt]
$H$ & Length of the control horizon \\
$k_0$ & Current time step at the start of the control horizon \\
$\III = \{k_0,\dots,k_0+H\}$ & Set of time steps over the control horizon \\
$\kappa_l \in [1, H+1]$ & Decision variable indicating \lane{l}'s traffic light switching time \\
$s_l(k) \in \{0,1\}$ & Traffic light state of \lane{l} at time $k$, $s_l(k)=0$ is red light and $s_l(k)=1$ is green light \\[2pt]
$\bar{k}_l$ & Time step of the previous switching event for \lane{l} \\
$\delta_{\min}$, $\delta_{\max}$ & Minimum and maximum allowable time gap between switches \\
$p_{l,i}(k)$ & Position of \CAV{l,i} at time step $k$ \\
$v_{l,i}(k)$ & Speed of \CAV{l,i} on lane $l$ at time step $k$ \\
$u_{l,i}(k)$ & Acceleration/deceleration input of \CAV{l,i} at time step $k$ \\[2pt]
$v_{\min}, v_{\max}$ & Minimum and maximum allowable speeds \\
$u_{\min}, u_{\max}$ & Minimum and maximum allowable accelerations \\[2pt]
\bottomrule
\end{tabular}
\end{table*}
In Table~\ref{tab:notation}, we summarize all variables and parameters used in the traffic light model, vehicle dynamics, and optimization formulation.
Next, we formulate a joint optimization problem for TLCs and CAVs and implement it in a receding horizon framework.
At each step, the controller solves an optimization problem to determine the best control actions, applies only the first control input, and then repeats the process at the next time step with updated system states. 
Therefore, receding horizon control can be robust against the uncertainty caused by HDVs.
Let $k_0$ be the current time step, $H$ be the control horizon, and $\III = \{k_0+1, \dots, k_0+H\}$ be the set of time steps in the next control horizon. 

In our formulation, each lane is allowed at most one traffic light switch over the control horizon.
Let $\kappa_{l} \in [1, H+1]$ so that the traffic light of \lane{l} switches at $k_0+ \left\lfloor \kappa_{l} \right\rfloor$ where $\left\lfloor \cdot \right\rfloor$ denotes the floor function.
If $\left\lfloor \kappa_{l} \right\rfloor = 1$, the traffic light switches at the next time step, while if $\left\lfloor \kappa_{l} \right\rfloor = H$, the traffic light switches at the last time step of the control horizon, and $\left\lfloor \kappa_{l} \right\rfloor = H+1$ means that the traffic light does not switch over the next control horizon.
At each time step $k$, let $s_l(k) \in \{0, 1\}$ be the traffic light state of \lane{l}, $l \in \LLL$, where $s_l(k) = 0$ if the traffic light is red and $s_l(k) = 1$ if the traffic light is green.
The traffic light states can be modeled as follows:
\begin{equation}
\label{eq:tlmodel}
s_l (k) =
\begin{cases}
1-s_l (k_0), & \text{if } k \ge k_0 + \kappa_{l}, \\
s_l (k_0), & \text{otherwise,}
\end{cases}     
\end{equation}
for all $k \in \III$. 
Next, we impose a constraint for the minimum and maximum time gaps between traffic light switches.
Let $\bar{k}_l$ be the time of the last traffic light switch of \lane{l}, $\delta_{\min} \in \RRplus$, and $\delta_{\max} \in \RRplus$ be the minimum and maximum time gaps, respectively. 
As a result, the next switching time must be within $[\delta_{\min}+\bar{k}_l, \delta_{\max}+\bar{k}_l]$.
Since we consider $\kappa_{l} \in [1, H+1]$, to guarantee feasibility for $\kappa_l$, the constraints are formulated as follows:
\begin{equation}
\label{eq:min-swt}
\begin{split}
\kappa_{l} &\ge \min \{ \delta_{\min}+\bar{k}_l-k_0, H+1\}, \\
\kappa_{l} &\le \max \{ \delta_{\max}+\bar{k}_l-k_0, 1\}.
\end{split}
\end{equation}
Note that the constraints \eqref{eq:min-swt} can be ignored if all vehicles in \lane{l} are CAVs.

For each \CAV{l,i}, let $p_{l,i}(k) \in \RR$, $v_{l,i}(k) \in \RR$, and $u_{l,i}(k) \in \RR$ be the position, speed, and control input (acceleration/deceleration) at time step $k$.
We consider the discrete double-integrator dynamics for \CAV{l,i} as follows:
\begin{equation}
\label{eq:integrator}
\begin{split}
p_{l,i}(k) &= p_{l,i}(k\!-\!1) \!+\! \Delta T v_{l,i}(k\!-\!1) \!+\! \frac{1}{2} \Delta T^2 u_{l,i}(k\!-\!1),  \\
v_{l,i}(k) &= v_{l,i}(k\!-\!1) \!+\! \Delta T u_{l,i}(k\!-\!1),
\end{split}
\end{equation}
for all $k \in \III$, where $\Delta T$ is the sample time. 
In addition, we impose the following speed and acceleration limit constraints,
\begin{equation}
\label{eq:bound-cons}
v_{\min} \le v_{l,i} (k) \le  v_{\max}, \; u_{\min} \le u_{l,i} (k-1) \le  u_{\max},
\end{equation}
for all $k \in \III$, where ${u_{\min}\in \RRminus}$ and ${u_{\max}\in \RRplus}$ are the minimum and maximum control inputs, ${v_{\min}\in \RRplus}$ and ${v_{\max}\in \RRplus}$ are the minimum and maximum speeds, respectively.

In this work, we assume that HDV trajectories over the control horizon are predicted using the constant acceleration model, meaning each HDV maintains a constant acceleration while adhering to the speed constraint in \eqref{eq:bound-cons}. 
The predicted trajectories, computed by the TLC of the lane, are then communicated to the CAVs to ensure rear-end safety constraints. 
While the constant acceleration model may not perfectly capture HDV behavior, the high-frequency receding horizon implementation can mitigate this limitation. 

\subsection{Coupling Safety Constraints}

Coordinating traffic lights and CAVs requires them to satisfy several coupling safety constraints.
First, the traffic lights must guarantee no lateral conflicts between a CAV and an HDV or between two HDVs.
Let \lane{l} and \lane{m} be two lanes with lateral conflicts, then the lights for those lanes cannot be both green if there are HDV-CAV or HDV-HDV conflicts, \ie
\begin{equation}
\label{eq:no-conf}
s_l(k) + s_m(k) \le 1,\; \forall k \in \III, \;\text{if}\; \eta(l, m, k_0) > 0,
\end{equation}
where $\eta (l,m,k_0)$ counts the number of CAV-HDV or HDV-HDV pairs that have lateral conflicts between \lane{l} and \lane{m} at the current time step $k_0$.
If a pair of vehicles travels on two intersecting lanes and neither vehicle has yet crossed the conflict point, they have lateral conflicts.
Next, we impose the following constraints, ensuring that the CAVs stop at red lights,
\begin{equation}
\label{eq:red-stop}
p_{l,i} (k) \le \psi_l,\; \forall k \in \III, \; \text{if} \; s_l(k) = 0.
\end{equation}
Note that we only need to impose \eqref{eq:red-stop} if \CAV{l,i} is the first vehicle in the queue of \lane{l}, \ie \CAV{l,i} has not crossed the stop line, and can stop by the line under maximum deceleration.
To avoid rear-end collisions, we consider safety constraints for each \CAV{l,i} if there is a preceding vehicle, which can be either a CAV or an HDV, as follows:
\begin{equation}
\label{eq:rearend}
p_{l,i} (k) + \tau v_{l,i} (k) + d_{\rm{min}} \le p_{l,i-1} (k),\; \forall k \in \III,
\end{equation}
where $\tau \in \RRplus$ is the desired time headway, and $d_{\rm{min}}\in \RRplus$ is the minimum distance.
Note that if \veh{l,i-1} is an HDV, then \eqref{eq:rearend} is considered as a local constraint of \CAV{l,i}, while \veh{l,i-1} is a CAV, \eqref{eq:rearend} is a coupling constraint between the two CAVs.
We also consider lateral safety constraints between two CAVs, \eg \CAV{l,i} and \CAV{m,j}, if they travel on two lanes with a lateral conflict point.
The lateral safety constraints can be formulated using OR statements, which require at every time step that two conflicting CAVs are not inside the conflict zone simultaneously, \ie
\begin{equation}
\begin{aligned}
\label{eq:lateral}
& p_{l,i} (k) - \psi_{l}  && \ge 0, \\
\text{OR}\quad & \phi_{l} - p_{l,i} (k)  && \ge 0, \\
\text{OR}\quad & p_{m,j} (k) - \psi_{m}  && \ge 0, \\
\text{OR}\quad & \phi_{m} - p_{m,j} (k)  && \ge 0.
\end{aligned}
\end{equation}

\subsection{Objective Functions}

In our formulation, each agent has its own separate objective.
For the TLCs, the goal is to maximize traffic throughput by prioritizing green lights for lanes with higher priority.
We define the following sigmoid-based priority function $\gamma_{l}$ for each lane that takes the vehicles' current positions as inputs
\begin{equation}
\gamma_{l} = \sum_{ \substack{i \in \CCC_l(k_0) \cup \HHH_l(k_0), \\ p_{l,i} (k_0) < \psi_l} } \mathrm{sigmoid} \bigg( \frac{p_{l,i} (k_0) - \psi_l/2}{\psi_l/2} \bigg).
\end{equation}
The idea of the above priority function is to take into account the number of vehicles in each lane while considering a higher priority for the vehicles near the intersection.

For the CAVs, we optimize the trajectories over the control horizon using a weighted sum of three terms: (1) maximizing the travel distance to reduce travel time, (2) tracking the desired speed, and (3) minimizing acceleration rates to reduce energy consumption.
Thus, our optimization problem minimizes the following objective
\begin{equation}
\label{eq:obj}
\begin{multlined}
J \Big(\{\bbsym{s}_{l}, \bbsym{p}_{l,i}, \bbsym{v}_{l,i} \bbsym{u}_{l,i}\}_{l \in \LLL, i \in \CCC_l(k_0)}\Big) = 
\sum_{k \in \III} \sum_{l \in \LLL} \Bigg( \!\! - \gamma_{l} \, s_l(k) \\ 
+ \sum_{i \in \CCC_l(k_0)} \!\! - \omega_p p_{l,i} (k) + \omega_v (v_{l,i} (k) - v_{\max})^2 \\
+ \omega_u u^2_{l,i} (k-1) \Bigg),
\end{multlined}
\end{equation}
where $\omega_p$, $\omega_v$ and $\omega_u \in \RRplus$ are the positive weights, while we let $\bbsym{s}_{l} = [s_{l}(k)]_{k \in \III}$, $\bbsym{p}_{l,i} = [p_{l,i}(k)]_{k \in \III}$, $\bbsym{v}_{l,i} = [v_{l,i}(k)]_{k \in \III}$, and $\bbsym{u}_{l,i} = [u_{l,i}(k-1)]_{k \in \III}$ be the vectors of variables for \TLC{l} and \CAV{l,i} over the control horizon, respectively.

In our problem formulation, the objective function is quadratic, while all the constraints are linear, which results in an MIQP problem.
The MIQP problem is given as follows:
\begin{equation}
\label{eq:main-prb}
\begin{split}
& \underset{\substack{\bbsym{s}_{l}, \bbsym{p}_{l,i}, \bbsym{v}_{l,i}, \bbsym{u}_{l,i} \\ \forall l \in \LLL, i \in \CCC_l(k_0)}}{\minimize} \;\;  J \Big(\{\bbsym{s}_{l}, \bbsym{p}_{l,i}, \bbsym{v}_{l,i}, \bbsym{u}_{l,i}\}_{l \in \LLL, i \in \CCC_l(k_0)}\Big), \\
& \subjectto 
\\
& \quad \eqref{eq:tlmodel}, \eqref{eq:min-swt}, \forall\, l \in \LLL, \\
& \quad \eqref{eq:integrator}, \eqref{eq:bound-cons}, \forall\, i \in \CCC_l(k_0), \\
& \quad \eqref{eq:no-conf}, \eqref{eq:red-stop}, \eqref{eq:rearend}, \eqref{eq:lateral},
\end{split}
\end{equation}
where the constraints hold for all time steps over the control horizon.

\subsection{Big-$M$ Constraint Formulation}

In mixed-integer optimization, continuous variables and binary variables are often used in conjunction with big-$M$ constraint formulations.
The general form of big-$M$ constraint formulation is given in the following definition.
\begin{definition}
Let $M \in \RRplus$ be a large positive number. A big-$M$ constraint involving a binary $\delta \in \{0,1\}$ and a function $\Phi(\cdot)$ of continuous variable $\bbsym{\nu}$ has the following form
\begin{equation}
\label{eq:bigM-def}
\Phi (\bbsym{\nu}) \le M \delta,
\end{equation} 
where $M$ must satisfy
\begin{equation}
\label{eq:validM}
M \ge \underset{\bbsym{\nu}}{\sup} \;\; \Phi (\bbsym{\nu}). 
\end{equation} 
The big-$M$ constraint \eqref{eq:bigM-def} implies that $\Phi (\bbsym{\nu}) \le 0$ is active if $\delta = 0$, otherwise, it is not activated.
Note that this definition also applies if $\delta$ is replaced by $(1-\delta)$.
\end{definition}

The big-$M$ constraints are commonly used to model OR statements or IF-ELSE statements as a set of linear constraints.
Using big-$M$ formulation, the IF-ELSE statement in \eqref{eq:tlmodel} is equivalent to 
\begin{equation}
\begin{split}
k - k_0 - \kappa_l & \ge -M (1 - s_{l} (k)), \\
k - k_0 - \kappa_l + \epsilon & \le M s_{l} (k), 
\end{split}
\end{equation}
if given $s_{l} (k_0) = 0$ and 
\begin{equation}
\begin{split}
k - k_0 - \kappa_l & \ge -M s_{l} (k), \\
k - k_0 - \kappa_l + \epsilon & \le M (1 - s_{l} (k)), 
\end{split}
\end{equation}
if given $s_{l} (k_0) = 1$,
where $\epsilon \in \RRplus$ is a sufficiently small number.
Next, we can formulate \eqref{eq:red-stop} as 
\begin{equation}
p_{l,i} (k) \le \psi_l + M s_l(k)
\end{equation}
The OR statement can be equivalently formulated using the following set of linear constraints using the big-$M$ method,
\begin{equation}
\label{eq:lateral-mi}
\begin{split}
p_{l,i} (k) - \psi_{l} & \ge - M c_{l,i} (k), \\
\phi_{l} - p_{l,i} (k) & \ge - M e_{l,i} (k), \\
p_{m,j} (k) - \psi_{m} & \ge - M c_{m,j} (k), \\
\phi_{m} - p_{m,j} (k) & \ge - M e_{m,j}(k), 
\end{split}
\end{equation}
and 
\begin{equation}
\label{eq:lateral-1}
c_{l,i} (k) + e_{l,i} (k) + c_{m,j} (k) + e_{m,j} (k) \le 3,
\end{equation}
for all $k \in \III$, where $c_{l,i}(k), e_{l,i}(k), c_{m,j}(k), e_{m,j}(k) \in \{0, 1\}$. 
The constraint \eqref{eq:lateral-1} implies that at least one binary variable among $c_{l,i} (k)$, $e_{l,i} (k)$, $c_{m,j} (k)$, and $e_{m,j} (k)$ must be $0$, which prevents two conflicting CAVs from occupying their conflict zones simultaneously.
Note that all the binary variables $c_{l,i} (k)$ and $e_{l,i}(k)$, $\forall i \in \CCC_l(k_0)$  are handled by \TLC{l}, leading to \eqref{eq:lateral-mi} is a coupling constraint between \TLC{l} and \CAV{l,i}, while \eqref{eq:lateral-1} is a coupling constraint between \TLC{l} and \TLC{m}. 
It can be observed that all the binary variables in the problem are involved in big-$M$ constraints.

\begin{remark}
Compared to the formulation presented in \citet{le2024distributed}, we do not consider fixed binaries $c_{l,i}, e_{l,i}, c_{m,j}, e_{m,j}$ used in the lateral safety constraints over the next control horizon.
As a result, it can better exploit the long-term coordination of CAVs within the receding horizon framework.
However, it significantly increases the number of binary variables in the MIQP.
\end{remark}

\section{Sequential Tightening of Constraint Coefficients}
\label{sec:tightening}

Constraint coefficient tightening is one of the techniques commonly used to reduce the number of explored nodes in the branch-and-bound method implemented in a presolve routine \citet{quirynen2023tailored}.
The main idea is to tighten the coefficients in the constraints involving integer variables so that solving the relaxed problem generates a solution that nearly satisfies the mixed-integer requirement.
As stated in \citet{quirynen2023tailored}, this technique is highly efficient for problems with big-$M$ constraints.
Inspired by this idea, we propose a heuristic strategy for constraint coefficient tightening sequentially, given the relaxed solution obtained at each iteration.

\subsection{Algorithm}

We first illustrate the method in a centralized algorithm to solve the MIQP problem presented in Section~\ref{sec:prb}.
Let $\bbsym{x} = [\bbsym{y}^\top, \bbsym{z}^\top]^\top$ be the vector of optimization variables, where $\bbsym{y} \in \YYY$ and $\bbsym{z} \in \ZZZ$ are the integer part and continuous part of $\bbsym{x}$, respectively.
Let us rewrite the optimization problem \eqref{eq:main-prb} in the following centralized form
\begin{equation}
\label{eq:miopt-cen}
\begin{split}
\underset{\bbsym{y} \in \YYY, \bbsym{z} \in \ZZZ}{\minimize} & \;\; f (\bbsym{y}, \bbsym{z}), \\
\text{subject to} & \;\; \bbsym{C} \bbsym{y} + \bbsym{D} \bbsym{z} \le \bbsym{b},
\end{split}
\end{equation}
where $\bbsym{C} \in \RR^{m\times n_y}$, $\bbsym{D} \in \RR^{m\times n_z}$, $\bbsym{b} \in \RR^m$.
We impose the following assumption on the feasibility of \eqref{eq:miopt-cen}.
\begin{assumption}
\label{assp:feasibility}
The problem \eqref{eq:miopt-cen} is feasible, \ie admits at least a solution.
\end{assumption}
Note that if Assumption~\ref{assp:feasibility} is not satisfied for our problem presented in Section~\ref{sec:prb}, we convert some hard constraints, \eg local constraints, into soft constraints by using the max penalty function (or equivalently, slack variables) and penalize the deviation in the objective function.

Next, we consider the following convex relaxation for \eqref{eq:miopt-cen}
\begin{equation}
\label{eq:miopt-cen-relax}
\begin{split}
\underset{\bbsym{y} \in \YYY', \bbsym{z} \in \ZZZ}{\minimize} & \;\; f (\bbsym{y}, \bbsym{z}), \\
\text{subject to} & \;\; \bbsym{C} \bbsym{y} + \bbsym{D} \bbsym{z} \le \bbsym{b},
\end{split}
\end{equation}
where $\YYY'$ is the set obtained from $\YYY$ by removing the integrality constraints. 
More specifically, for our problem, we replace the constraint set $\{0, 1\}$ by $[0,1]$ for all the binary variables.

The main idea is to reconstruct a solution of \eqref{eq:miopt-cen} starting from a solution of \eqref{eq:miopt-cen-relax} by sequentially tightening the big-$M$ constraints.
Since $\bbsym{C}$ and $\bbsym{b}$ may involve $M$, we can express the linear constraint in \eqref{eq:miopt-cen} in the following form
\begin{equation}
\label{eq:cen_con}
\bbsym{C}(M) \bbsym{y} + \bbsym{D} \bbsym{z} \le \bbsym{b}(M).
\end{equation}
Note that $\bbsym{D}$ does not involve $M$.
Let $\III_M$ be the set of the indices of big-$M$ constraints, and without loss of generality, we assume that the first $m_M$ constraints of \eqref{eq:cen_con} are big-$M$ constraints.
Let us consider the $q$-th constraint that involves big-$M$, $q \in \III_M$ in the following form
\begin{equation}
\label{eq:con_with_M}
\bbsym{C}_{[q]} (M) \bbsym{y} + \bbsym{D}_{[q]} \bbsym{z} \le b_{[q]} (M),
\end{equation}
where $\bbsym{C}_{[q]}$ and $\bbsym{D}_{[q]}$ are the $q$-th row of $\bbsym{C}$ and $\bbsym{D}$, respectively.
To update the coefficients of \eqref{eq:con_with_M}, we first identify the element in $\bbsym{C}_{[q]}$ that involves $M$, \eg $r$-th element, and let us denote it by $\bbsym{C}_{[q,r]}$.
\hilitediff{
To update each coefficient $\bbsym{C}_{[q,r]}$, we restrict the corresponding big-$M$ coefficient to a finite grid.
For each $q\in\III_M$, let
\begin{equation}
\label{eq:M-grid}
\MMM_{[q]} =
\left\{ M_{[q]}^{0}, \alpha M_{[q]}^{0}, \alpha^2 M_{[q]}^{0}, \ldots, M_{[q]}^{\min} \right\},
\; 0<\alpha<1,
\end{equation}
where $M_{[q]}^{0}$ is chosen as a valid big-$M$ value satisfying \eqref{eq:validM}, and $M_{[q]}^{\min}>0$ is a prescribed lower bound.
The grid is constructed so that $M_{[q]}^{\min}$ is the smallest allowable value in $\MMM_{[q]}$.
At iteration $t$, after solving the relaxed problem, let $\bar{\bbsym{y}}^{(t)}$ denote the relaxed binary vector, and let $\bar y_{[r]}^{(t)}$ be the relaxed binary variable associated with the $q$-th big-$M$ term.
If $\bar y_{[r]}^{(t)}$ is already binary, the corresponding big-$M$ coefficient is unchanged.
Otherwise, we first form the candidate tightened coefficient
\begin{equation}
\label{eq:M-candidate}
\widetilde M_{[q]}^{(t+1)}
=
M_{[q]}^{(t)}
\max\left\{\xi,\bar y_{[r]}^{(t)}\right\},
\end{equation}
where $\xi>0$ is a small constant used to avoid overly aggressive tightening when $\bar y_{[r]}^{(t)}$ is close to zero.
We then project this candidate onto the finite grid of admissible big-$M$ values:
\begin{equation}
\label{eq:M-update}
M_{[q]}^{(t+1)} =
\begin{cases}
M_{[q]}^{(t)}, & \bar y_{[r]}^{(t)}\in\{0,1\},\\[1mm]
\Pi_{[q]}^{-}\!\left(\widetilde M_{[q]}^{(t+1)}\right), & \text{otherwise},
\end{cases}
\end{equation}
where
\begin{equation}
\label{eq:M-proj}
\Pi_{[q]}^{-}(a) := \max\left\{ M\in\MMM_{[q]}: M\leq M_{[q]}^{(t)},\; M\leq a
\right\}.
\end{equation}
If the set in \eqref{eq:M-proj} is empty, we set $\Pi_{[q]}^{-}(a)=M_{[q]}^{\min}$.
The updated value $M_{[q]}^{(t+1)}$ is then used to update the corresponding entry $\bbsym C_{[q,r]}$, yielding the new coefficient $\bbsym C_{[q,r]}^{(t+1)}$.
In other words, we first compute a candidate tightened big-$M$ coefficient from the relaxed binary value, project this candidate onto the finite grid of admissible big-$M$ values, and then update the corresponding constraint coefficient.
The proposed finite-grid rule has two useful properties.
First, the sequence $\{M_{[q]}^{(t)}\}_{t\ge 0}$ is monotone non-increasing and bounded below by $M_{[q]}^{\min}$.
Second, because $\MMM_{[q]}$ contains only finitely many values, each sequence $\{M_{[q]}^{(t)}\}_{t\ge 0}$ stabilizes after finitely many updates.
Therefore, since $\III_M$ is finite, there exists a finite iteration index $T_M$ such that
\begin{equation}
\label{eq:M-finite-stabilization}
M_{[q]}^{(t)} = M_{[q]}^{(T_M)}, \; \forall q\in \III_M,\; \forall t\ge T_M .
\end{equation}
After the tightening coefficients stabilize, the constraint coefficients remain fixed, and the subsequent iterations are applied to a fixed tightened convex relaxation.
}
Note that the relaxed problem at iteration $t$ can be infeasible if the constraints are overly tightened.
To address this issue, we utilize the max penalty function to find the solution with minimal violation, resulting in the following optimization problem at each iteration:
\begin{equation}
\label{eq:miopt-cen-slack}
\begin{split}
\underset{\bbsym{y} \in \YYY', \bbsym{z} \in \ZZZ}{\minimize} & \;\; f (\bbsym{y}, \bbsym{z}) + \!\! \sum_{q \in \III_M} \max(0, \bbsym{C}_{[q]}^{(t)} \bbsym{y} + \bbsym{D}_{[q]} \bbsym{z} - b_{[q]}), \\
\text{subject to} & \;\; \bbsym{C}_{[q]} \bbsym{y} + \bbsym{D}_{[q]} \bbsym{z} \le b_{[q]}, \; q \notin \III_M. \\
\end{split}
\end{equation}

\hilitediff{The centralized algorithm with sequential tightening of the constraint coefficients is thus given in Algorithm~\ref{algo:Alg1}.
The relaxed binary variables obtained from Algorithm~\ref{algo:Alg1} are used to identify high-confidence binary decisions and construct the reduced MIQP recovery problem described in the next section.
}

\subsection{\hilitediff{Confidence-Guided Binary Recovery}}
\label{subsec:confidence_binary_recovery}

\hilitediff{Once Algorithm~\ref{algo:Alg1} converges and returns relaxed binary variables in $[0,1]$, we recover the binary solutions.
A common recovery step is to round all relaxed binaries independently.
However, rounding variables that are close to $0.5$ may introduce infeasibility. 
To improve constraint satisfaction, we fix only high-confidence binary variables and leave uncertain variables to be re-optimized.}

\hilitediff{
Let $(\bar{\bbsym{y}}, \bar{\bbsym{z}})$ be the relaxed solution returned by Algorithm~\ref{algo:Alg1}, where $\bar{\bbsym{y}}$ contains the relaxed binary variables. 
For each binary variable $\bar y_j$, we define the confidence score as follows:
\begin{equation}
\chi_j=\left|\bar y_j-\frac{1}{2}\right|.
\end{equation}
Given a threshold $\chi_{\min}\in[0,1/2]$, we define the set of high-confidence binaries as
\begin{equation}
\BBB_{\mathrm{fix}} =
\{j \in \BBB:\chi_j\geq \chi_{\min}\},
\end{equation}
where $\BBB$ is the index set of binary variables. For
$j \in \BBB_{\mathrm{fix}}$, we fix
\begin{equation}
\hat y_j =
\begin{cases}
1, & \bar y_j\geq 1/2,\\
0, & \bar y_j<1/2.
\end{cases}
\end{equation}
The remaining low-confidence binaries are retained as binary variables and
optimized in a reduced MIQP problem as:
\begin{equation}
\begin{aligned}
\minimize_{\bbsym{y}, \bbsym{z}, \bbsym{s}} \quad
& f(\bbsym{y}, \bbsym{z}) + \bbsym{\lambda}^\top \bbsym{s} \\
\subjectto \quad
& \bbsym{C} \bbsym{y} + \bbsym{D} \bbsym{z} \leq \bbsym{b} + \bbsym{s},\\
& y_j=\hat y_j,\quad j\in\BBB_{\mathrm{fix}},\\
& y_j\in\{0,1\},\quad j\in\BBB\setminus\BBB_{\mathrm{fix}},\\
& \bbsym{s} \geq 0.
\end{aligned}
\label{eq:confidence_recovery_miqp}
\end{equation}
The slack variables $\bbsym{s}$ quantify remaining constraint violations, with larger penalties assigned to safety-related constraints. 
By fixing high-confidence binaries, the recovery step branches only over low-confidence binaries, yielding a reduced MIQP that is substantially smaller than the original problem.}

\begin{algorithm}[tb!]
\caption{\hilitediff{Centralized Sequential-Tightening Algorithm}}
\label{algo:Alg1}
\hilitediff{
\begin{algorithmic}[1]  
\Require $t_{\mathrm{max}}$, $\epsilon$, finite sets $\{\MMM_{[q]}\}_{q\in\III_M}$
\State Initialize big-$M$ coefficients $\{M_{[q]}^{(0)}\}_{q\in\III_M}$
\For{$t = 0,1,\dots,t_{\mathrm{max}}-1$}
\State Solve the relaxed QP problem \eqref{eq:miopt-cen-slack} to obtain $(\bar{\bbsym y}^{(t+1)},\bar{\bbsym z}^{(t+1)})$
\State Update the big-$M$ coefficients using the finite-grid rule \eqref{eq:M-update}
\If{$M_{[q]}^{(t+1)} = M_{[q]}^{(t)}$, $\forall q\in \III_M$} %
    \State \textbf{break}
\EndIf
\EndFor
\State Solve the final relaxed QP problem \eqref{eq:miopt-cen-slack} to obtain $(\bar{\bbsym y},\bar{\bbsym z})$
\State \Return $(\bar{\bbsym y},\bar{\bbsym z})$
\end{algorithmic}
}
\end{algorithm}

\subsection{Illustrative Example}

\begin{figure}
\centering
\includegraphics[width=0.48\textwidth]{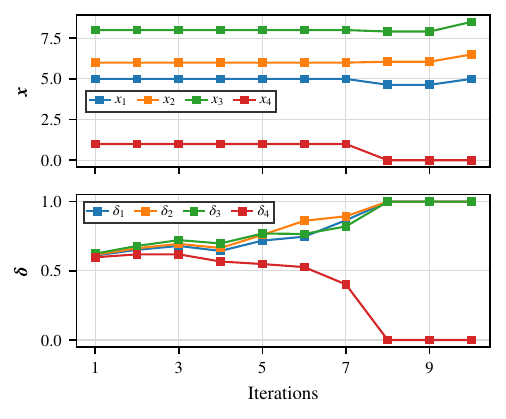}
\caption{The solutions obtained from the relaxed QP of \eqref{eq:example}, for continuous variables (top), and integer variables (bottom).}
\label{fig:example}
\end{figure}

To better illustrate the sequential constraint-tightening process, we consider the following simple example.
\begin{example}
\label{example}
\begin{equation}
\label{eq:example}
\begin{split}
\underset{\bbsym{x}}{\minimize} & \;\; \bbsym{x}^\top \bbsym{x} + \bbsym{q}^\top \bbsym{x}, \\
\text{subject to} & \;\; \bbsym{x} \le [5,12,9,6]^\top,\; \bbsym{x} - M \, \bbsym{\delta} \le \bbsym{0}, \\
& \;\; \bbsym{1}^\top \bbsym{x} \le 20, \; \bbsym{1}^\top \bbsym{\delta} \le 3, \\
& \;\; \bbsym{x} \in \RR^4, \; \bbsym{\delta} \in \{0,1\}^4,
\end{split}
\end{equation}
where $\bbsym{q}^\top = [-30,-20,-24,-10]$.
\end{example} 
The mixed-integer optimal solution of \eqref{eq:example} is $\bbsym{x} = [5.0, 6.5, 8.5, 0.0]^\top$, $\bbsym{\delta} = [1, 1, 1, 0]^\top$.
The parameters of proximal ADMM are chosen as: $\rho = 0.1$, $\beta = 0.5$, $\gamma = 1$. 
Using our algorithm, convergence is achieved after $10$ iterations.
In Fig.~\ref{fig:example}, we show the solution trajectory obtained from solving the relaxed QP over iterations, which demonstrates that the relaxed binary variables converge.

\section{Distributed Mixed-Integer Quadratic Programming}
\label{sec:admm}

In this section, we present the distributed algorithm to solve the MIQP problem based on proximal ADMM \citep{deng2017parallel} and the constraint tightening technique.
Though ADMM algorithms were originally introduced as a tool for convex optimization problems, they have proven to be a powerful heuristic method even for NP-hard nonconvex problems \citep{boyd2011distributed}.
A natural extension of the algorithm presented in the last section is to utilize proximal ADMM until convergence in the inner loop to find the solution of the relaxed QP at each iteration of Algorithm~\ref{algo:Alg1}.
However, we propose a single-loop algorithm where, at each iteration, we perform proximal ADMM while tightening the constraints.

\subsection{Proximal-ADMM Relaxation with Sequential Tightening}
\label{subsec:admm_tightening}

First, we describe the communication between the agents in our multi-agent optimization problem in the following definition. 
\begin{definition}
Let $N$ be the total number of agents, $\VVV = \{1, \dots, N\}$ be the set of agents, and $\EEE \subset \VVV \times \VVV$ be the set of edges.
We assume that the edges are undirected.
For each \agent{i}, $i \in \VVV$, let $\NNN_i = \{j \in \VVV \;|\; (i,j) \in \EEE\}$ be the set of its neighbors.
\end{definition}
Next, we rewrite the optimization problem \eqref{eq:main-prb} in the following general form with separable objectives and coupling constraints
\begin{equation}
\label{eq:miopt-prb}
\begin{split}
\underset{\bbsym{x}_i \in \XXX_i}{\minimize} & \;\; \sum_{i=1}^N \; f_i (\bbsym{x}_i), \\
\text{subject to} & \;\; \bbsym{A}_{i} \bbsym{x}_i \le \bbsym{b}_i, \\
& \;\; \sum_{i=1}^N \bbsym{C}_{i} \bbsym{x}_i \le \bbsym{d}, 
\end{split}
\end{equation}
where $\XXX_i$ is the mixed-integer-valued set for $\bbsym{x}_i$,
$f_i$ is the local objective function of each \agent{i}, $\bbsym{A}_{i}$, and $\bbsym{b}_i$ are the matrices and vectors of coefficients for the local constraints, while $\bbsym{C}_{i}$ and $\bbsym{d}$ are the matrices and vectors of coefficients for the coupling constraints, respectively.
The optimization problem \eqref{eq:miopt-prb} follows the standard formulation in \citet{deng2017parallel}, which covers the case where the coupling constraints are pairwise in our problem.

We convert an inequality constraint into an equality constraint by introducing auxiliary allocation variables.
Let $\bbsym{w}_{i}$ be the allocation variables \agent{i} which satisfy $\sum_{i=1}^{N} \bbsym{w}_{i} = \bbsym{d}$.
Thus, we can formulate the equivalent problem \eqref{eq:miopt-prb} as follows:
\begin{subequations}
\label{eq:miopt-prb-dist}
\begin{align}
\underset{\bbsym{x}_i \in \XXX_i, \bbsym{w}_{i}}{\minimize} & \;\; \sum_{i=1}^N \; f_i (\bbsym{x}_i), \\
\text{subject to}
& \;\; \bbsym{A}_{i} \bbsym{x}_i \le \bbsym{b}_i, \\
& \;\; \bbsym{C}_{i} \bbsym{x}_i \le \bbsym{w}_{i}, \\
& \;\; \sum_{i=1}^{N} \bbsym{w}_{i} = \bbsym{d},
\end{align}
\end{subequations}
where let $\bbsym{w} = [\bbsym{w}_{i}]^\top_{i \in \VVV}$ is the concatenated vector of all auxiliary variables.

The convex relaxation of \eqref{eq:miopt-prb-dist} is given by 
\begin{subequations}
\label{eq:qp-prb-dist}
\begin{align}
\underset{\bbsym{x}_i \in \XXX_i', \bbsym{w}_{i}}{\minimize} & \;\; \sum_{i=1}^N \; f_i (\bbsym{x}_i), \\
\text{subject to}
& \;\; \bbsym{A}_{i} \bbsym{x}_i \le \bbsym{b}_i, \\
& \;\; \bbsym{C}_{i} \bbsym{x}_i \le \bbsym{w}_{i}, \\
& \;\; \sum_{i=1}^N \bbsym{w}_{i} = \bbsym{d}. \end{align}
\end{subequations}

The augmented Lagrangian for the problem with equality constraints is formulated as follows:
\begin{equation}
\begin{multlined}
\LLL (\bbsym{x}, \bbsym{w}, \bbsym{\lambda}) 
= \sum_{i=1}^N f_i (\bbsym{x}_i) 
+ \Big< \bbsym{\lambda}, \sum_{i=1}^N \bbsym{w}_i - \bbsym{d} \Big> \\
+ \frac{\rho}{2} \norm{\sum_{i=1}^N \bbsym{w}_i - \bbsym{d}}^2,
\end{multlined}
\end{equation}
where $\bbsym{\lambda}$ is the vector of dual variables (Lagrangian multipliers), and $\rho > 0$ are positive constants.

The proximal ADMM algorithm for solving \eqref{eq:qp-prb-dist} works as follows.
At each iteration $t+1$, each \agent{i} solves the following $x$-minimization problem in parallel
\begin{equation}
\label{eq:x_min}
\begin{split}
\bar{\bbsym{x}}_i^{(t+1)}, \bbsym{w}_i^{(t+1)} = &\underset{\bbsym{x}_i \in \XXX_i', \bbsym{w}_i}{\argmin} \; f_i(\bbsym{x}_i) + \frac{\beta}{2} \norm{\bbsym{w}_i - \bbsym{w}_i^{(t)}}^2  \\
& \hspace{-4mm} + \frac{\rho}{2} \norm{\bbsym{w}_i - \big( \bbsym{d} - \sum_{j=1,j \neq i}^{N} \bbsym{w}_j^{(t)} \big) + \frac{\bbsym{\lambda}^{(t)}}{\rho}}^2, \\
\subjectto
& \;\; \bbsym{A}_{i} \bbsym{x}_i \le \bbsym{b}_i, \\
& \;\; \bbsym{C}_{i} \bbsym{x}_i \le \bbsym{w}_{i},
\end{split}
\end{equation} 
where $\beta > 0$.
Given the relaxed solution $\bar{\bbsym{x}}_i^{(t+1)}$, \agent{i} adjusts the constraint coefficients based on the big-$M$ update strategy presented in \eqref{eq:M-update}.
Next, \agent{i} transmits $\bbsym{w}_i^{(t+1)}$ to \agent{j}, $j \in \NNN_i$.
After receiving all the information from the neighbors,  \agent{i} updates the dual variables as follows:
\begin{equation}
\label{eq:dual}
\bbsym{\lambda}^{(t+1)} = \bbsym{\lambda}^{(t)} + \gamma \rho\, \Big(\sum_{i=1}^{N} \bbsym{w}_i^{(t+1)} - \bbsym{d}\Big),
\end{equation}
where $\gamma > 0$ is a damping coefficient. 
Note that in a distributed implementation, each agent can keep a copy of the dual variables $\bbsym{\lambda}$ and update them in parallel.
Compared to regular ADMM, proximal ADMM adds a proximal term $\frac{\beta}{2} \norm{\bbsym{w}_i - \bbsym{w}_i^{(t)}}^2$ to the $x$-minimization problem and a damping coefficient to the dual update. 
The proximal ADMM algorithm can fully exploit parallel computation for solving subproblems and achieve convergence at a rate of $\OOO(1/k)$.
The algorithm with distributed implementation can also be summarized in Algorithm~\ref{algo:Alg2}.
Similar to the centralized algorithm, in the second stage of the distributed algorithm, we fix the binary variables found in the first stage and utilize proximal ADMM to find the optimal values for continuous variables with the original big-$M$ constraints.
Since the second stage follows the standard implementation of proximal ADMM, the details are omitted.
In practice, the second stage converges very quickly since the solution returned from the first stage is close to the true solution of the QP problem with fixed binary variables.

\begin{figure}
\centering
\includegraphics[width=0.48\textwidth]{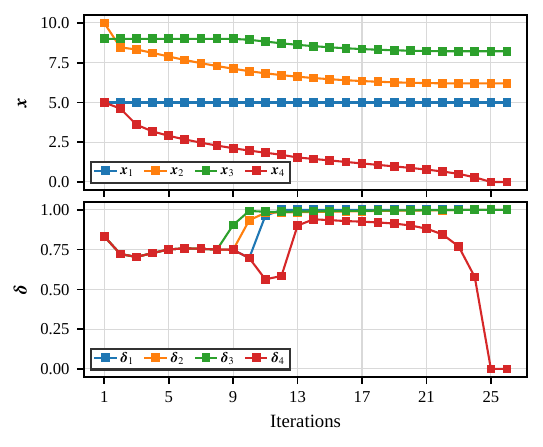}
\caption{The solutions obtained from the relaxed QP of \eqref{eq:example}, for continuous variables (top), and integer variables (bottom).}
\label{fig:example_2}
\end{figure}

To better demonstrate the convergence property of the distributed algorithm, we revisit Example~\ref{example} presented in Section~\ref{sec:tightening}, in which we consider 4 agents coordinating to solve the problem.
The variable of each \agent{i} is $\bbsym{x}_{[i]}$ and $\bbsym{\delta}_{[i]}$.
The agents share two coupling constraints $\bbsym{1}^\top \bbsym{x} \le 20$ and $\bbsym{1}^\top \bbsym{\delta} \le 3$ as shown in \eqref{eq:example}.
The parameters of proximal ADMM are chosen as: $\rho = 0.1$,  $\beta = 0.5$, and $\gamma = 1$.
Our numerical simulation shows that the distributed algorithm converges after $26$ iterations, and its solution trajectory over iterations is shown in Fig.~\ref{fig:example_2}.

\begin{algorithm}[tb!]
\caption{\hilitediff{Distributed Sequential-Tightening Algorithm}}
\label{algo:Alg2}
\hilitediff{
\begin{algorithmic}[1]
\Require $t_{\mathrm{max}}$, $\epsilon$, $\rho$, $\beta$, $\gamma$, finite sets $\{\mathcal M_{[q]}\}_{q\in\III_M}$
\State Initialize big-$M$ coefficients $\{M_{[q]}^{(0)}\}_{q\in\III_M}$
\State Initialize primal variables $\{\bar{\bbsym x}_i^{(0)},\bbsym w_i^{(0)}\}_{i\in\mathcal V}$ and dual variable $\bbsym\lambda^{(0)}$
\For{$t=0,1,\dots,t_{\mathrm{max}}-1$}
\State Each \agent{i} solves \eqref{eq:x_min} in parallel to obtain $\bar{\bbsym x}_i^{(t+1)}$ and $\bbsym w_i^{(t+1)}$
\State Each \agent{i} updates its associated big-$M$ coefficients using \eqref{eq:M-update}
\State Each \agent{i} transmits $\bbsym w_i^{(t+1)}$ to \agent{j}, $j\in\mathcal N_i$
\State Each \agent{i} updates the dual variable using \eqref{eq:dual}
\If{$M_{[q]}^{(t+1)}=M_{[q]}^{(t)}$, $\forall q\in\III_M$}
    \State \textbf{break}
    \EndIf
\EndFor
\State Solve the final tightened convex relaxation with fixed big-$M$ coefficients using proximal ADMM to obtain $(\bar{\bbsym x},\bar{\bbsym w})$.
\State \Return relaxed solution $\bar{\bbsym x}=(\bar{\bbsym y},\bar{\bbsym z})$
\end{algorithmic}
}
\end{algorithm}

\subsection{\hilitediff{Distributed Implementation of Binary Recovery}}
\label{subsec:bb-admm-recovery}

\hilitediff{
Similar to the confidence-guided binary recovery for the centralized algorithm, after Algorithm~\ref{algo:Alg2} converges, we recover the low-confidence binary variables by solving a reduced MIQP. 
The reduced MIQP is handled by a branch-and-bound procedure, while each node relaxation is solved using the same distributed proximal-ADMM procedure as Algorithm~\ref{algo:Alg2}. 
At a branch-and-bound node $\nu$, let $\BBB_0^\nu$ and $\BBB_1^\nu$ be the sets of variables fixed to $0$ and $1$, respectively. 
The node relaxation is
\begin{align}
\min_{\bbsym{y},\bbsym{z},\bbsym{s}}\quad
& f(\bbsym{y},\bbsym{z})
+\frac{\mu}{2}\|\bbsym{z}-\bar{\bbsym{z}}\|^2
+\lambda^\top\bbsym{s}
\label{eq:bb-node}\\
\text{s.t.}\quad
& \bbsym{C}\bbsym{y}+\bbsym{D}\bbsym{z}
\le \bbsym{b}+\bbsym{s}, \nonumber\\
& y_j=\hat y_j,\; j\in\BBB^{\rm fix}, \nonumber\\
& y_j=0,\; j\in\BBB_0^\nu, \;
  y_j=1,\; j\in\BBB_1^\nu, \nonumber\\
& 0\le y_j\le 1,\;
j\in\BBB^{\rm rec}\setminus
(\BBB_0^\nu\cup\BBB_1^\nu), \nonumber\\
& \bbsym{s}\ge 0. \nonumber
\end{align}
For every node $\nu$, \eqref{eq:bb-node} is a convex QP and can therefore be solved by proximal ADMM in a distributed manner. 
The branch-and-bound tree is coordinated by the central coordinator, which maintains the incumbent solution and applies standard pruning rules. 
Thus, the recovery step is distributed at the level of node relaxations, while the branch-and-bound search is coordinated centrally.
If the branch-and-bound tree is fully explored and each node relaxation is solved to optimality, this recovery step returns an optimal solution of the reduced MIQP induced by fixing \(\BBB^{\rm fix}\). 
}

\section{\hilitediff{Theoretical Properties}}
\label{sec:theo}

\hilitediff{In this section, we analyze the theoretical properties of the proposed distributed algorithm. 
We first show in Proposition~\ref{proposition:1} that the finite-grid tightening rule guarantees finite-time stabilization of the big-$M$ coefficients. 
Next, we show in Proposition~\ref{proposition:2} that, after the tightening coefficients stabilize, Algorithm~\ref{algo:Alg2} reduces to proximal ADMM applied to a fixed tightened convex QP relaxation, and therefore converges under the standard proximal-ADMM parameter condition.}

\hilitediff{\begin{proposition}
\label{proposition:1}
For each big-$M$ constraint $q\in\III_M$, let $\{M_{[q]}^{(t)}\}_{t\ge 0}$ be generated by \eqref{eq:M-update}.
Then $\{M_{[q]}^{(t)}\}_{t\ge 0}$ is monotone non-increasing and stabilizes after finite iterations. 
That is, there exists a finite index $T_q<\infty$ and a value $M_{[q]}^\star \in \MMM_{[q]}$ such that
\begin{equation}
M_{[q]}^{(t)}=M_{[q]}^\star,
\; \forall t\ge T_q .
\end{equation}
Moreover, since $\III_M$ is finite, there exists $T_M<\infty$ such that
\begin{equation}
M_{[q]}^{(t)}=M_{[q]}^\star,
\;
\forall q\in\III_M,\; \forall t\ge T_M .
\end{equation}
\end{proposition}}

\hilitediff{\begin{proof}
For each $q\in\III_M$, the update rule either keeps $M_{[q]}^{(t)}$ unchanged or maps it to a smaller value on the set $\MMM_{[q]}$.
Therefore, $M_{[q]}^{(t+1)}\le M_{[q]}^{(t)}, \; \forall t\ge 0$.
Moreover, since $M_{[q]}^{(t)}\in\MMM_{[q]}$ for all $t$ and $M_{[q]}^{\min}$ is the smallest grid value, the sequence is bounded below.
Because $\MMM_{[q]}$ contains finitely many elements, a monotone sequence taking values in $\MMM_{[q]}$ can decrease only finitely many times.
Thus, there exists a finite index $T_q$ such that
$M_{[q]}^{(t)}=M_{[q]}^\star,\; \forall t\ge T_q$,
for some $M_{[q]}^\star\in\MMM_{[q]}$.
Finally, since the number of big-$M$ constraints is finite, we can define
\[T_M:=\max_{q\in\III_M} T_q < \infty.\]
Hence, all big-$M$ coefficients stabilize after $T_M$ iterations.
The proof is thus complete.
\end{proof}}

\hilitediff{\begin{proposition}
\label{proposition:2}
By Proposition~\ref{proposition:1}, under the finite-grid update rule, there exists a finite index \(T_M\) such that
\begin{equation}
M_{[q]}^{(t)} = M_{[q]}^{(T_M)}, \; \forall q\in\III_M,\quad \forall t\ge T_M .
\end{equation}
Consequently, for all $t\ge T_M$, Algorithm~2 reduces to proximal ADMM applied to a fixed tightened convex QP relaxation.
If this fixed relaxation is feasible and the ADMM parameters satisfy
\begin{equation}
\label{eq:admm_params}
\beta > \rho\left(\frac{N}{2-\gamma}-1\right),
\end{equation}
then the proximal ADMM iterates converge to a solution of the fixed tightened convex relaxation.
\end{proposition}}

\hilitediff{
\begin{proof}
Since the number of big-$M$ constraints is finite, there exists a finite index $T_M$ such that
$M_{[q]}^{(t)} = M_{[q]}^\star, \; \forall q\in\III_M,\; \forall t\ge T_M$.
Therefore, after iteration $T_M$, all big-$M$ coefficients remain fixed.
Hence, the local and coupling constraint sets in the relaxed problem no longer change, and Algorithm~2 becomes proximal ADMM applied to a fixed convex QP relaxation.
If this fixed tightened relaxation is feasible, then the standard convergence result for parallel multi-block proximal ADMM \citep{deng2017parallel} applies under the condition \eqref{eq:admm_params}.
Thus, the iterates generated by Algorithm~\ref{algo:Alg2} converge to a solution of the fixed tightened convex relaxation.
\end{proof}}

\hilitediff{
Proposition~\ref{proposition:2} guarantees convergence of the distributed algorithm after the finite-grid tightening coefficients stabilize.
Specifically, Proposition~\ref{proposition:2} shows that the finite-grid update makes all big-$M$ coefficients stabilize after finitely many iterations.
Once this stabilization occurs, the feasible set no longer changes, and Algorithm~\ref{algo:Alg2} reduces to proximal ADMM applied to a fixed tightened convex QP relaxation.
Under the feasibility and parameter conditions in Proposition~\ref{proposition:2}, the subsequent iterates converge to a solution of this fixed relaxation.
}

\hilitediff{
\begin{remark}
The final tightened relaxation is still a relaxation of the original mixed-integer problem, and therefore, this convergence result does not imply global optimality or exact feasibility for the original MIQP. Accordingly, we present the proposed method as a practically effective distributed heuristic for the MIQP, followed by a confidence-guided binary recovery step to improve the feasibility of the final mixed-integer solution.
\end{remark}
}

\section{Numerical Studies}
\label{sec:sim}

In this section, we demonstrate the control performance of the proposed framework by numerical simulations.

\subsection{Simulation Setups}

We validated our framework using a mixed-traffic simulation environment in \texttt{SUMO} interfacing with the Julia programming language via \texttt{TraCI} \citep{lopez2018microscopic} and the \texttt{PyCall} package.
\hilitediff{We considered an intersection with a \SI{300}{m}-long control zone for each lane, where the conflict zone is located between \SI{190}{m} and \SI{210}{m} from the control-zone entry.}
We conducted simulations for three traffic volumes: $1200$, $1400$, and $1600$ vehicles per hour, and six different penetration rates: $0\%$, $20\%$, $40\%$, $60\%$, $80\%$, and $100\%$.
In the simulations, the vehicles are randomly assigned to the lanes, with balanced entry rates across four directions.
The parameters of the problem formulation are given in Table~\ref{tab:parameters}.

\begin{table}[t!]
\caption{Parameters of the problem formulation.}
\label{tab:parameters} 
\centering
\begin{tabular}{ K{0.1\textwidth} | K{0.08\textwidth} | K{0.1\textwidth} | K{0.08\textwidth} }
\textbf{Parameters} & \textbf{Values} & \textbf{Parameters} & \textbf{Values} \\
\midrule[0.5pt] %
$H$ & $20$ & $\Delta T$ & \SI{0.5}{s} \\
$\delta_{\rm{min}}$ & $20$ & $\delta_{\rm{max}}$ & $100$ \\
$v_\mathrm{max}$ & $\SI{15.0} {m/s}$ & $v_\mathrm{min}$ & $\SI{0.0} {m/s}$ \\
$a_\mathrm{max}$ & $\SI{3.0} {m/s^2}$ & $a_\mathrm{min}$ & $\SI{-4.0} {m/s^2}$ \\
$\tau$ & $\SI{1.0}{s}$ & $d_{\rm{min}}$ & $\SI{6.0}{m}$ \\
$\omega_p$ & $10^{0}$ & $\omega_v$ & $10^{0}$ \\ 
$\omega_u$ & $10^{-1}$ & $M$ & $10^3$ \\
\end{tabular}
\end{table}

\subsection{Simulation Results}

We first evaluate the performance of our proposed optimization algorithm by comparing the solutions with the \texttt{GUROBI} solver (see \citet{gurobi}).
We randomly generate problems with different numbers of agents and initial conditions.
In Table~\ref{tab:accu}, we evaluate the proposed algorithm using three metrics: (1) accuracy rates, (2) optimality gaps, and (3) constraint violations.
For each metric, we report the average over $1,000$ problem instances for each fixed number of agents.
\hilitediff{We compute the accuracy rates by comparing the binary solution from our algorithm with the optimal binary solution returned by \texttt{GUROBI}.
The optimality gaps are computed as the difference between the approximate and optimal objective values returned by \texttt{GUROBI}, normalized by the absolute optimal objective value.
The constraint violations are computed as the total violation of all constraints in each problem instance over the predicted solution horizon.
Therefore, the zero constraint-violation values reported after confidence-guided binary recovery indicate that the recovered solution of the reduced MIQP satisfies the constraints for the tested problem instances.
Note that these values should not be interpreted as a formal closed-loop safety guarantee.
In the receding-horizon implementation, only the first control input is applied at each time step, and the optimization problem is solved again using updated vehicle states and predictions.}
\hilitediff{As shown in Table~\ref{tab:accu}, before confidence-guided binary recovery, the proposed distributed algorithm achieves binary accuracy above $95\%$ for all tested problem sizes, with optimality gaps generally below $2\%$.
However, the recovered relaxed solutions may still exhibit constraint violations due to inaccuracies in the binary decisions.
After applying the confidence-guided binary recovery step, the solution quality improves substantially: the accuracy increases to above $98\%$, the optimality gaps become nearly zero, and the constraint violations of the recovered reduced-MIQP solutions are eliminated in all tested cases.
These results indicate that the proposed recovery step effectively improves feasibility of the recovered mixed-integer solutions while preserving high solution quality.}
Regarding the computational aspect, our distributed algorithm can find an approximate solution in a reasonable amount of time, and it is more scalable with the number of agents than the centralized approach, as shown in Table~\ref{tab:time}.

\begin{table*}[!t] 
\centering 
\caption{\hilitediff{Comparison of solution quality before and after confidence-guided binary recovery for Algorithm~\ref{algo:Alg2}.}} \vspace{2mm}
\label{tab:accu} 
\hilitediff{
\begin{tabular}{ K{3.0cm} | K{3.0cm} | K{1.5cm} | K{1.5cm} | K{1.5cm} | K{1.5cm} } 
\toprule[1.0pt] 
Metric & Stage & {$7$} & {$12$} & {$17$} & {$22$} \\ 
\midrule[0.5pt] 
\multirow{2}{*}{Accuracy rate ($\%$)} 
& Before recovery & $97.9$ & $96.8$ & $96.3$ & $95.7$ \\ 
& After recovery  & $98.3$ & $99.6$ & $99.8$ & $99.9$ \\ 
\midrule[0.5pt] 
\multirow{2}{*}{Optimality gap ($\%$)} 
& Before recovery & $0.70$ & $1.17$ & $1.85$ & $0.95$ \\ 
& After recovery  & $0.06$ & $0.00$ & $0.00$ & $0.00$ \\ 
\midrule[0.5pt] 
\multirow{2}{*}{Constraint violation} 
& Before recovery & $8.14$ & $26.87$ & $40.74$ & $54.34$ \\ 
& After recovery  & $0.00$ & $0.00$ & $0.00$ & $0.00$ \\ 
\bottomrule[1.0pt] 
\end{tabular} 
}
\end{table*}

\begin{table}[!t]
\centering
\caption{Average computation time (in seconds) for the centralized optimization using \texttt{GUROBI} solver and the proposed distributed optimization algorithm (Algorithm~\ref{algo:Alg2}) across different numbers of agents.} \vspace{2mm}
\label{tab:time}
\begin{tabular}{ K{2.9cm} | K{0.8cm} | K{0.8cm} | K{0.8cm} | K{0.8cm}  }
\toprule[1.0pt]
Number of agents & {$15$} & {$20$} & {$25$} & {$30$} \\ 
\midrule[0.5pt]
Algorithm~\ref{algo:Alg2}'s computation time
& 0.06
& 0.12
& 0.20
& 0.27 \\
GUROBI's computation time
& 0.04 
& 0.13
& 0.26 
& 0.53 \\
\bottomrule[1.0pt]
\end{tabular}
\end{table}

\begin{figure*}
\centering
\begin{subfigure}{0.48\textwidth}
\includegraphics[scale=0.42]{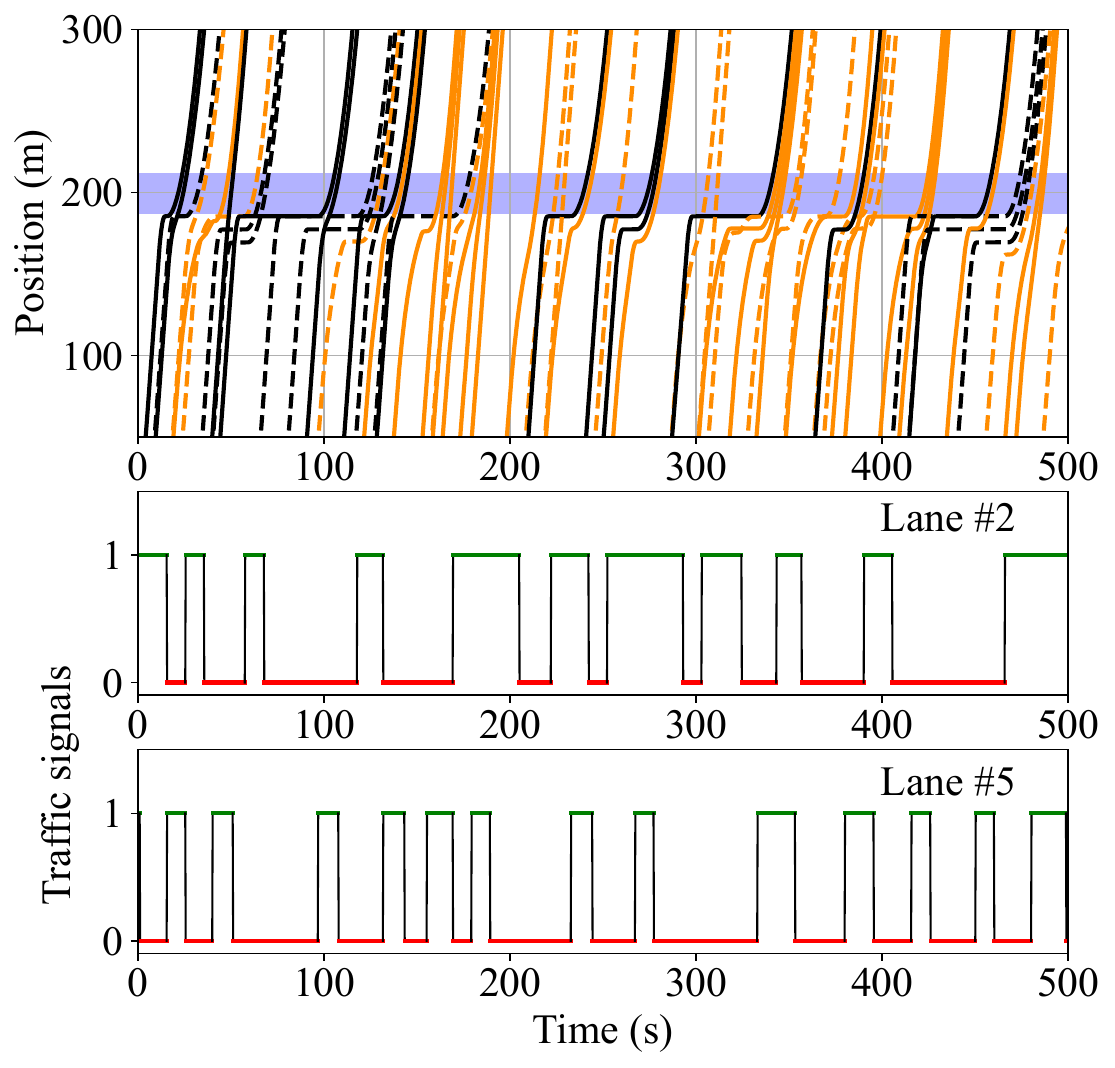}  
\caption{$60 \%$ penetration rate}
\end{subfigure}
\begin{subfigure}{0.48\textwidth}
\includegraphics[scale=0.42]{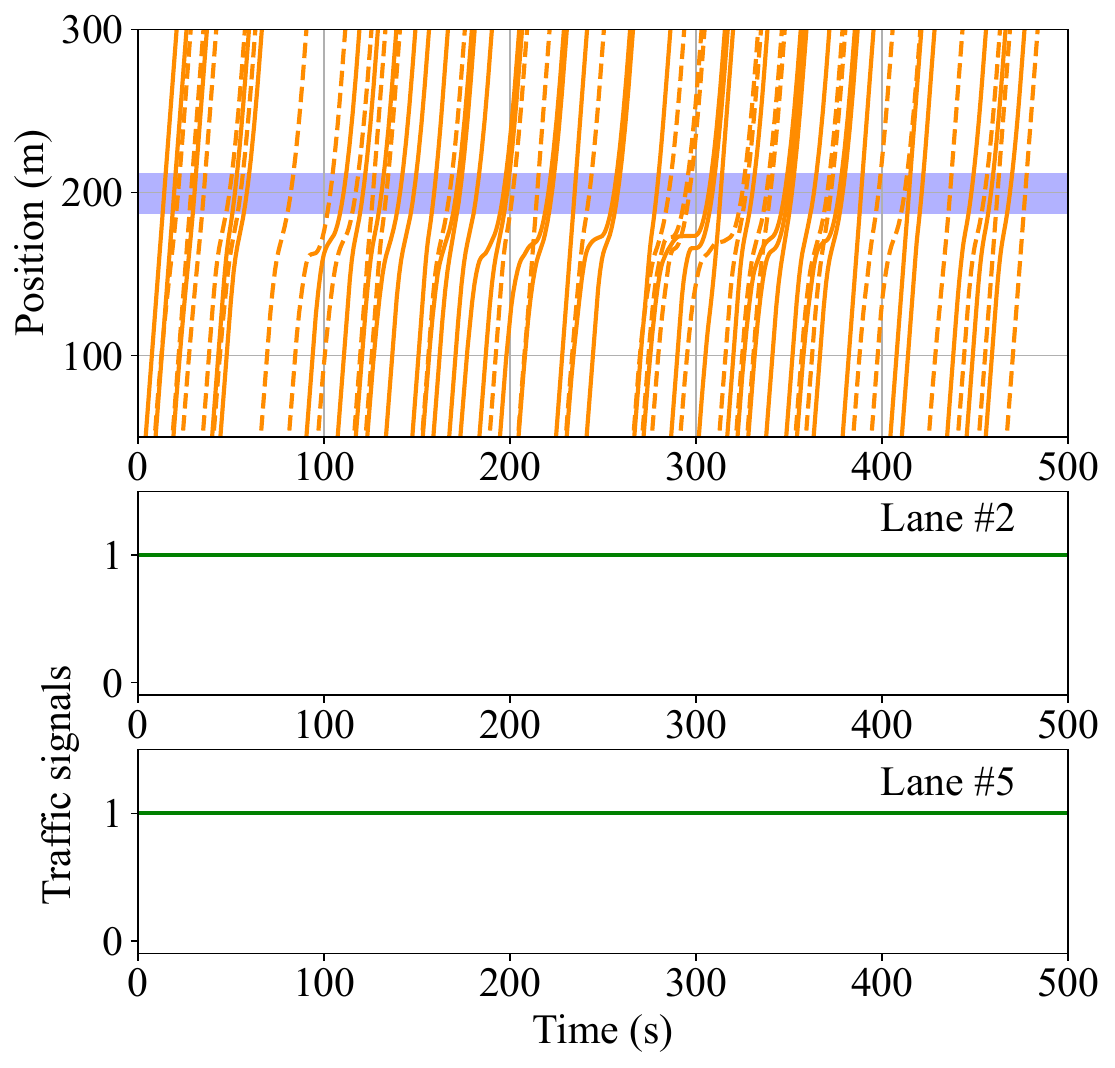}
\caption{$100 \%$ penetration rate}
\end{subfigure}
\caption{Vehicle trajectories and traffic light states for two conflicting lanes. The trajectories for CAVs and HDVs are represented by orange and black curves, respectively. 
Dashed and solid curves distinguish the vehicles moving on \lane{2} and \lane{5}, respectively. The conflict zone is depicted as a shaded blue area.
} 
\label{fig:traj}
\vspace{-3pt}
\end{figure*}

Videos and data from the simulations can be found at \url{https://sites.google.com/view/tlc-cav-admm}.
The simulations show that the proposed framework efficiently manages traffic light systems and CAVs under different traffic volumes and penetration rates. It enables CAVs to coordinate intersection crossings while reducing complete stops at traffic lights, particularly when penetration rates are high, such as $80\%$ or $100\%$.
In Fig.~\ref{fig:traj}, we show an example of the vehicle trajectories and traffic light states for two conflicting lanes in simulations with $60\%$ and $100\%$ penetration rates and $1600$ vehicles per hour.

\begin{figure}[!tb]
\centering
\includegraphics[scale=0.35]{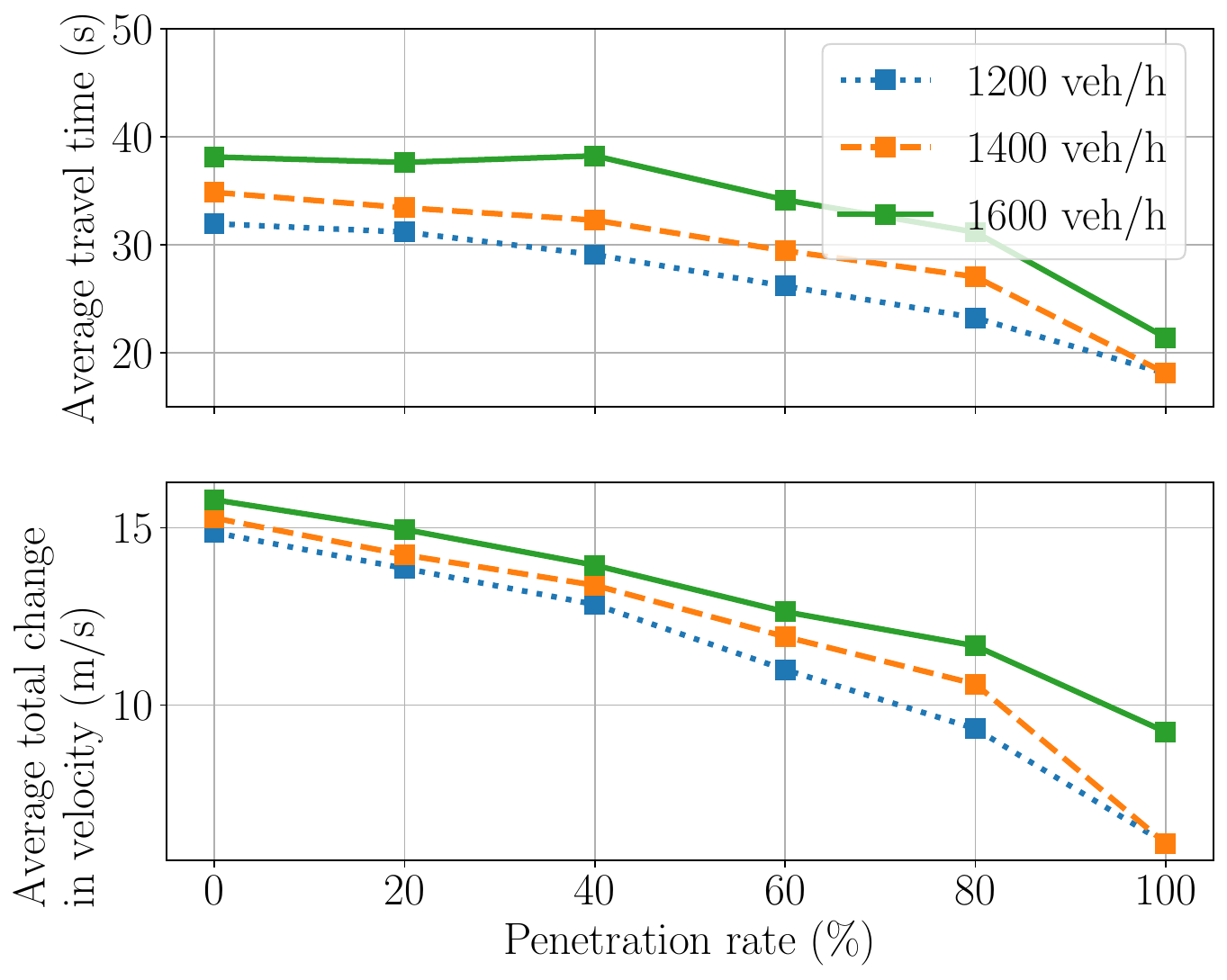}
\caption{Average travel time (top) and average total change in velocity (bottom) under different penetration rates and traffic volumes.}
\label{fig:metr}
\vspace{-1mm}
\end{figure}

\begin{figure}
\centering
\includegraphics[scale=0.35]{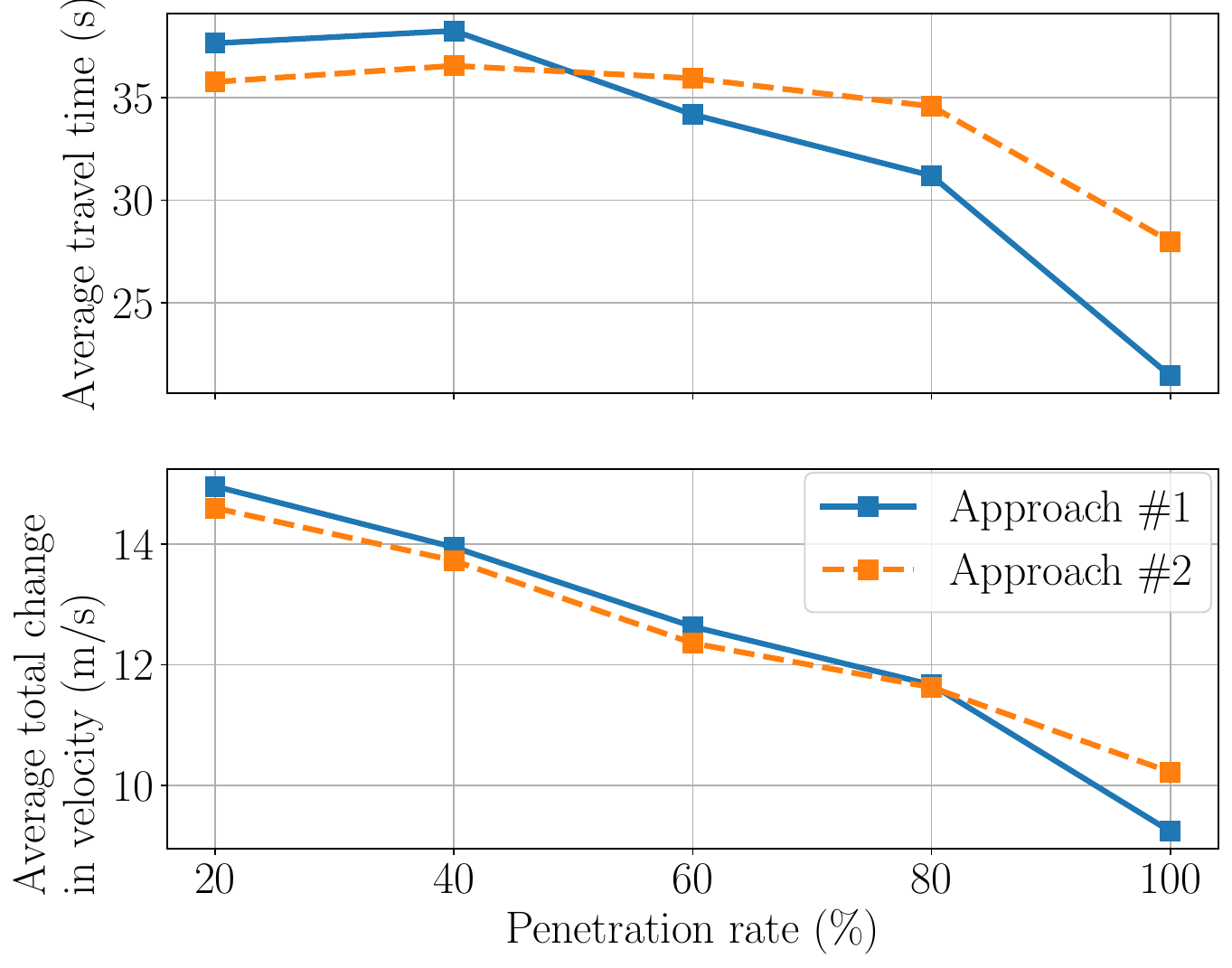}
\caption{Comparison of average travel time (top) and average total change in velocity (bottom) between two formulations: this work ($\#1$) and \citet{le2024distributed} ($\#2$).}
\label{fig:baseline}
\vspace{-3pt}
\end{figure}

To thoroughly evaluate the performance of our proposed framework, we run simulations of 30 minutes for different traffic volumes and penetration rates to collect statistical results.
\hilitediff{We compare two metrics: (1) the average travel time and (2) the total change in velocity averaged over all vehicles. 
The total change in velocity of each vehicle while traveling in the control zone is computed as the accumulated magnitude of velocity changes over time, and therefore has units of $m/s$.}
The results are shown in Fig.~\ref{fig:metr}.
Overall, in all three examined traffic volumes, starting from a penetration rate of $60\%$, the framework achieves remarkable travel time improvement compared to the scenario with pure HDVs.
The framework also performs well at lower penetration rates ($20\%$ and $40\%$), and differences compared to the pure HDV case remain relatively small.
Moreover, we can observe that CAV penetration generally leads to lower total change in velocity, which may result in less energy consumption (see \citet{malikopoulos2008optimal}), and a more comfortable travel experience. 
Especially under $100\%$ penetration, the CAV coordination can significantly improve the metrics.
That is because the framework allows CAVs to coordinate their intersection crossings while mitigating full stops at traffic lights. 

To demonstrate the improvement obtained by fully leveraging CAV coordination for intersection crossing, we compare the proposed formulation with the baseline formulation developed in our prior work \citep{le2024distributed}.
The statistics are computed from simulations conducted under the high traffic volume of $1600$ vehicles per hour and are shown in Fig.~\ref{fig:baseline}.
The top figure shows that Approach~\#1, proposed in this paper, results in a significant reduction in average travel time as the CAV penetration rate increases, whereas Approach~\#2 exhibits a more gradual decrease.
This suggests that the proposed CAV coordination formulation improves traffic efficiency at higher penetration rates compared with the formulation in \citet{le2024distributed}.
The bottom figure shows that the total change in velocity is generally comparable between the two approaches.
\hilitediff{At lower penetration rates, Approach~\#1 may result in a slightly larger total change in velocity than Approach~\#2, indicating more active acceleration/deceleration behavior. 
This is likely because the additional coordination flexibility provided by the time-varying lateral crossing-order binaries cannot be fully exploited under low CAV penetration rates. 
In this range, the differences between the two approaches remain relatively small: the travel-time difference is less than about $5$--$6\%$, and the total-change-in-velocity difference is less than about $3$--$4\%$. 
As the CAV penetration rate increases, the uncertainty introduced by HDVs decreases, and a larger fraction of vehicles can be coordinated, allowing Approach~\#1 to more effectively exploit its additional flexibility.
}

\section{Conclusions}
\label{sec:conc}

In this paper, we presented an MIQP framework designed to coordinate traffic signals and CAVs within mixed-traffic intersections, effectively leveraging the benefits of long-term CAV coordination. 
To address the challenges posed by the resulting multi-agent MIQP, characterized by a substantial number of binary variables, we developed a distributed optimization algorithm capable of solving the problem within a reasonable timeframe. 
This distributed optimization approach utilized proximal ADMM for resolving the convex relaxation and employed a sequential constraint tightening heuristic to uphold integrality constraints. 
The simulation results demonstrated that our distributed algorithm successfully improved the performance of the control framework across various CAV penetration rates and traffic volumes, with notable improvements observed as the CAV penetration rate increased. 
We believe that further enhancements to the algorithm can be achieved through additional techniques aimed at optimizing mixed-integer solutions and minimizing computation time, which will be the focus of our future research.

\bibliographystyle{abbrvnat}
\bibliography{references,refs_IDS}

\par
\begin{wrapfigure}{l}{0.10\textwidth}
\centering
\includegraphics[width=0.13\textwidth]{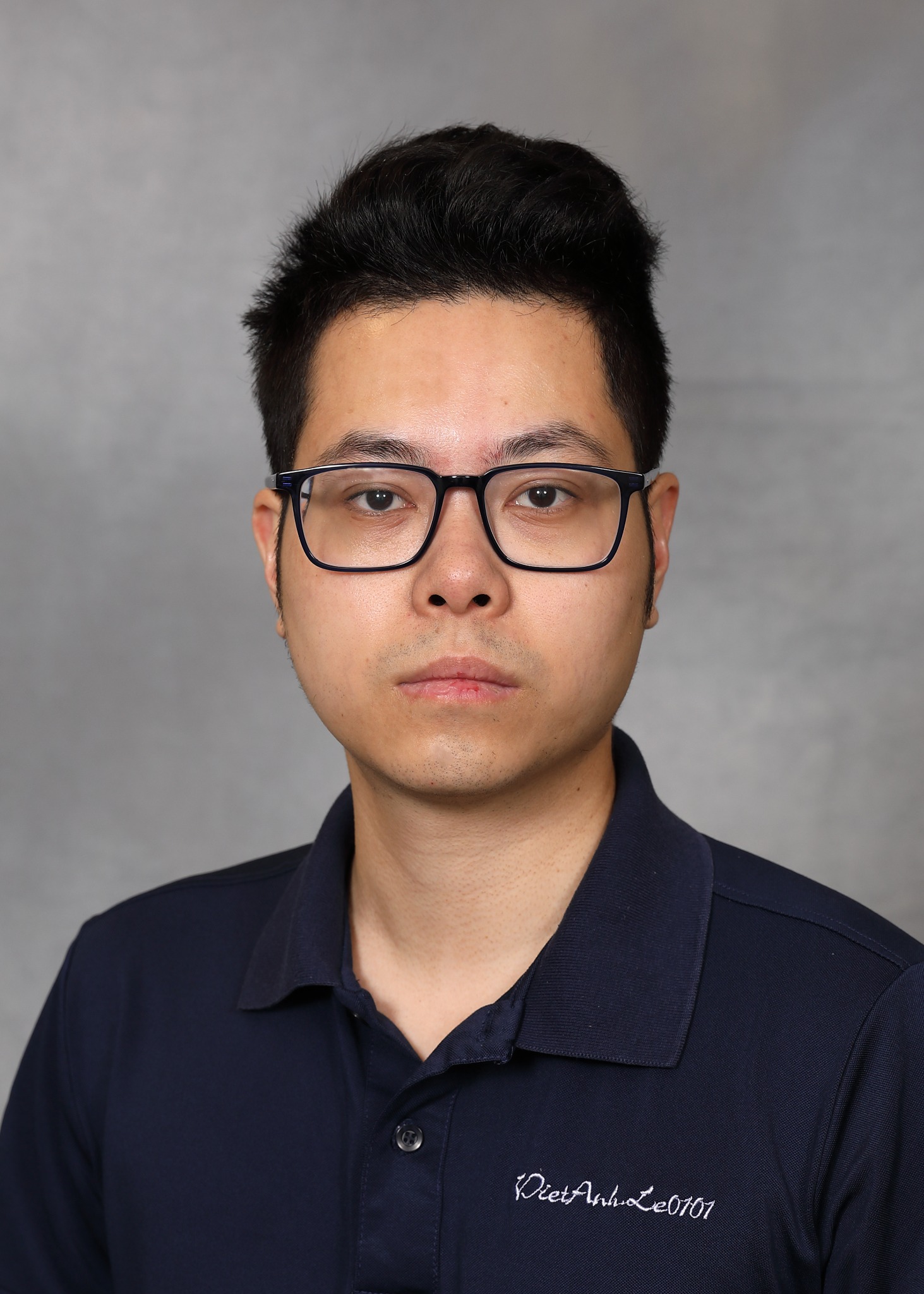}
\end{wrapfigure}
\textbf{Viet-Anh Le} is a postdoctoral researcher in the Department of Electrical and Systems Engineering at the University of Pennsylvania, USA. 
He received his Ph.D. in Mechanical Engineering from the University of Delaware in 2025, his M.Sc. in Informatics from Northern Arizona University in 2021, and his B.Sc. in Automation and Control Engineering (Talented Program) from Hanoi University of Science and Technology, Vietnam, in 2019. 
From September 2023 to August 2025, he was a visiting graduate scholar in Systems Engineering at Cornell University, and from June to August 2023, he was a research intern at Honda Research Institute USA. 
Dr. Le received the IEEE Intelligent Transportation Systems Society (ITSS) Best Dissertation Award (2nd Place) and was selected as a Penn AI Fellow in 2026. 
His research interests include scientific machine learning for optimization and control, model predictive control, distributed optimization, and autonomous and multi-agent systems.

\par
\begin{wrapfigure}{l}{0.10\textwidth}
\centering
\includegraphics[width=0.13\textwidth]{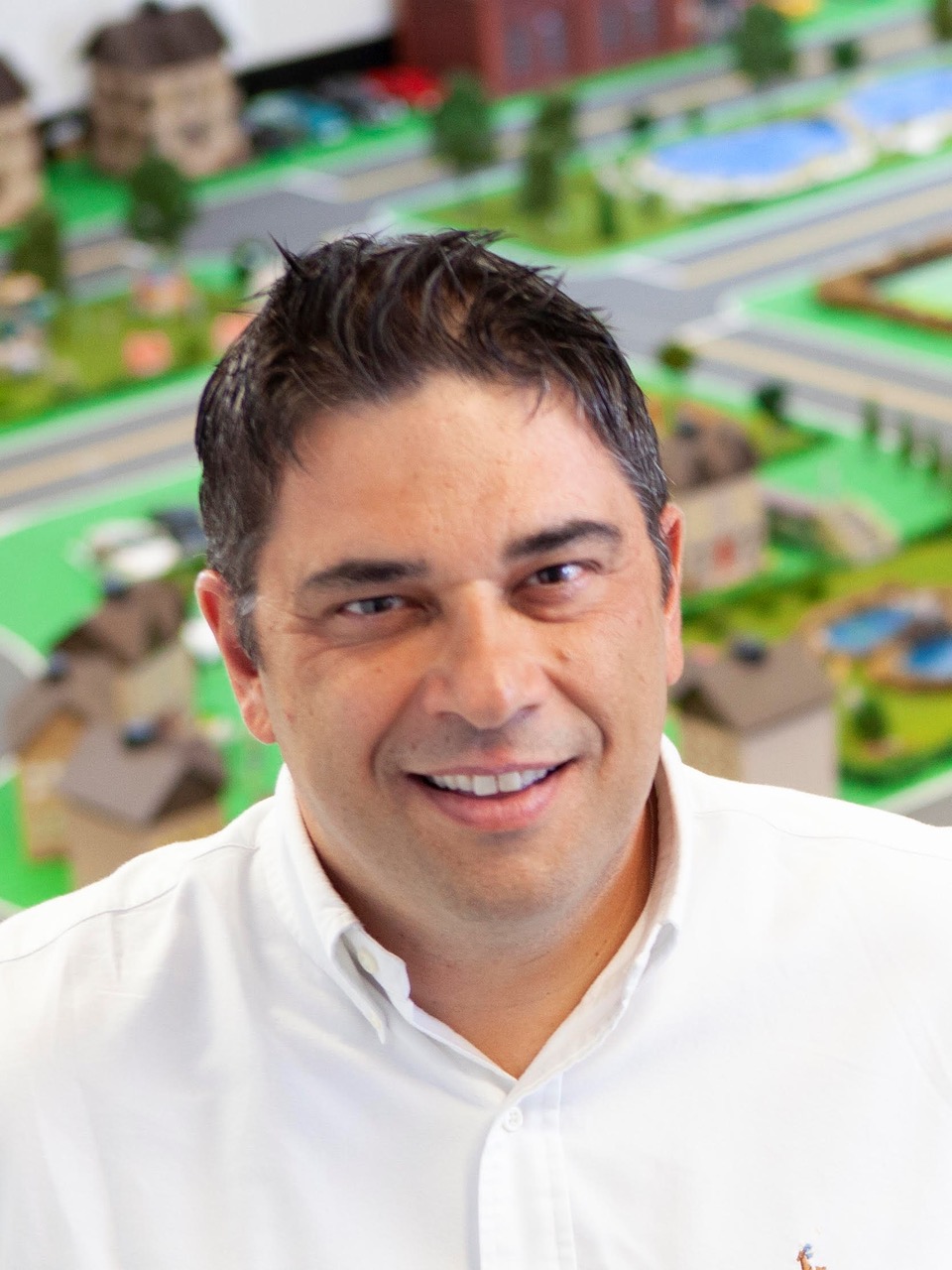}
\end{wrapfigure}
\textbf{Andreas A. Malikopoulos} received the Diploma in mechanical engineering from the National Technical University of Athens, Greece, in 2000, and the M.S. and Ph.D. degrees in mechanical engineering from the University of Michigan, Ann Arbor, MI, USA, in 2004 and 2008, respectively.

He is a Professor in the School of Civil and Environmental Engineering at Cornell University, Ithaca, NY, USA, and the Director of the Information and Decision Science (IDS) Laboratory. He holds courtesy appointments and affiliations with Applied Mathematics, Systems Engineering, Cornell Robotics, Electrical and Computer Engineering, and Mechanical Engineering, and is a Faculty Fellow of the Cornell Atkinson Center for Sustainability. Prior to these appointments, he was the Terri Connor Kelly and John Kelly Career Development Professor in the Department of Mechanical Engineering (2017--2023) and the founding Director of the Sociotechnical Systems Center (2019--2023) at the University of Delaware (UD). Before he joined UD, he was the Alvin M. Weinberg Fellow (2010--2017) in the Energy \& Transportation Science Division at Oak Ridge National Laboratory (ORNL), the Deputy Director of the Urban Dynamics Institute (2014--2017) at ORNL, and a Senior Researcher in General Motors Global Research \& Development (2008--2010). His research interests include stochastic control, team theory, decentralized decision-making, optimization, and the intersection of learning and control, with applications to multi-agent systems, robotics, and emerging mobility systems.

Dr. Malikopoulos received the IEEE Intelligent Transportation Systems (ITS) Society Young Researcher Award in 2019, the IEEE ITS Society Outstanding Application Award in 2026, and Cornell University's Faculty Excellence in Mentoring Postdoctoral Scholars Award in 2026. His research has been recognized with several best paper and student paper awards. He was named a National Academy of Sciences Kavli Frontiers of Science Scholar in 2012. He serves as an Associate Editor of \textit{Automatica} and \textit{IEEE Transactions on Automatic Control}, and as a Senior Editor of \textit{IEEE Transactions on Intelligent Transportation Systems}. He is a member of the IEEE ITS Society Board of Governors and an IEEE ITS Society Distinguished Lecturer for 2024--2027. He is a Fellow of ASME and a member of SIAM, AAAS, and Sigma Xi.

\end{document}

%% file: defs.tex
\newcommand{\probP}{\text{I\kern-0.15em P}}
\newcommand{\CAV}[1]{CAV\textendash\ensuremath{(#1)}\xspace}

\newcommand{\veh}[1]{vehicle\textendash\ensuremath{(#1)}\xspace}
\newcommand{\lane}[1]{lane\textendash\ensuremath{#1}\xspace}

\newcommand{\TLC}[1]{TLC\textendash\ensuremath{#1}\xspace}

\newcommand{\agent}[1]{agent\textendash\ensuremath{#1}\xspace}

\newcommand{\bbsym}[1]{\ensuremath{\boldsymbol{#1}}}